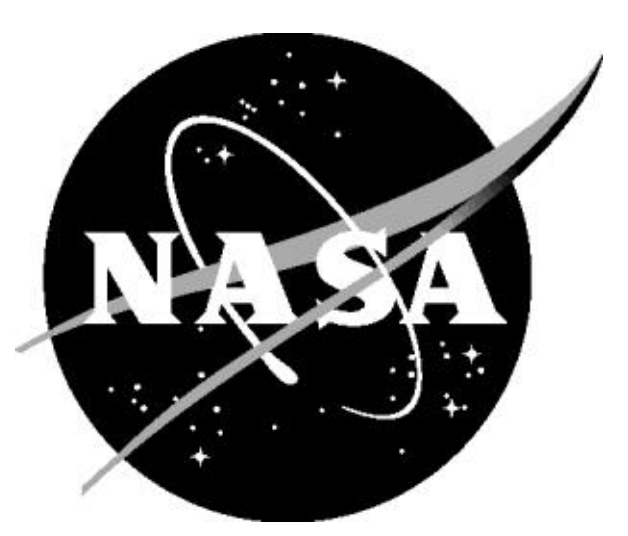

# Mechanisms of Electrostatic Charge Formation and Retention in Lunar Regolith

*Minhyeok Kim and Hyun Jung Kim*
*Korea Advanced Institute of Science and Technology, Daejeon, South Korea*

*Sang H. Choi*
*ELA*
*NASA Langley Research Center*

*Ashley Daeun Jung*
*Imperial College London, London, United Kingdom*

## NASA STI Program Report Series

Since its founding, NASA has been dedicated to the advancement of aeronautics and space science. The NASA scientific and technical information (STI) program plays a key part in helping NASA maintain this important role.

The NASA STI program operates under the auspices of the Agency Chief Information Officer. It collects, organizes, provides for archiving, and disseminates NASA's STI. The NASA STI program provides access to the NTRS Registered and its public interface, the NASA Technical Reports Server, thus providing one of the largest collections of aeronautical and space science STI in the world. Results are published in both non-NASA channels and by NASA in the NASA STI Report Series, which includes the following report types:

- TECHNICAL PUBLICATION. Reports of completed research or a major significant phase of research that present the results of NASA Programs and include extensive data or theoretical analysis. Includes compilations of significant scientific and technical data and information deemed to be of continuing reference value. NASA counterpart of peer-reviewed formal professional papers but has less stringent limitations on manuscript length and extent of graphic presentations.

- TECHNICAL MEMORANDUM. Scientific and technical findings that are preliminary or of specialized interest, e.g., quick release reports, working papers, and bibliographies that contain minimal annotation. Does not contain extensive analysis.

- CONTRACTOR REPORT. Scientific and technical findings by NASA-sponsored contractors and grantees.

- CONFERENCE PUBLICATION. Collected papers from scientific and technical conferences, symposia, seminars, or other meetings sponsored or co-sponsored by NASA.

- SPECIAL PUBLICATION. Scientific, technical, or historical information from NASA programs, projects, and missions, often concerned with subjects having substantial public interest.

- TECHNICAL TRANSLATION. English-language translations of foreign scientific and technical material pertinent to NASA's mission.

Specialized services also include organizing and publishing research results, distributing specialized research announcements and feeds, providing information desk and personal search support, and enabling data exchange services.

For more information about the NASA STI program, see the following:

- Access the NASA STI program home page at http://www.sti.nasa.gov

- Help desk contact information:

https://www.sti.nasa.gov/sti-contact-form/ and select the "General" help request type.

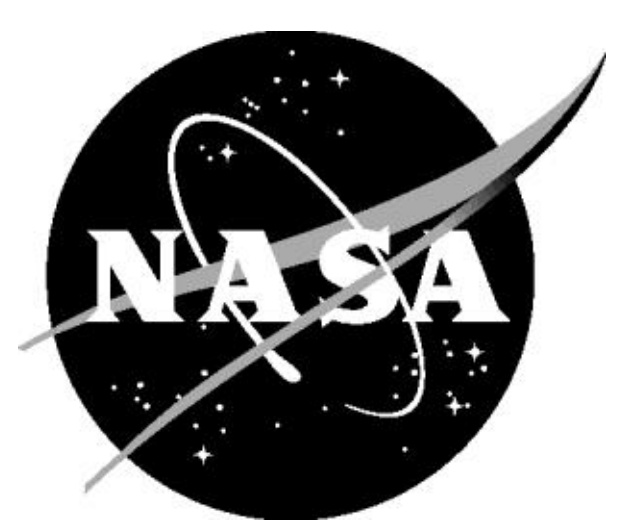

# Mechanisms of Electrostatic Charge Formation and Retention in Lunar Regolith

*Minhyeok Kim and Hyun Jung Kim*
*Korea Advanced Institute of Science and Technology, Daejeon, South Korea*

*Sang H. Choi*
*ELA*
*NASA Langley Research Center*

*Ashley Daeun Jung*
*Imperial College London, London, United Kingdom*

# Table of Contents

## Acronyms

| | |
|---|---|
| ACE | Advanced Composition Explorer |
| ARTEMIS | Acceleration, Reconnection, Turbulence and Electrodynamics of the Moon's Interaction with the Sun |
| CPD | Contact Potential Difference |
| CPLEE | Charged Particle Lunar Environment Experiment |
| CRaTER | Cosmic Ray Telescope for the Effects of Radiation |
| CSDA | Continuous Slowing Down Approximation |
| EFI | Electric Field Instrument |
| EPAM | Electron, Proton, and Alpha Monitor |
| ER | Electron Reflectometer |
| ESA | Electrostatic Analyzer |
| ESA-S1 | Electron Spectrum Analyzer |
| EUV | Extreme Ultraviolet |
| FUV | Far Ultraviolet |
| GCR | Galactic Cosmic Ray |
| ICME | Interplanetary Coronal Mass Ejection |
| IEA | Ion Energy Analyzer |
| LP | Lunar Prospector |
| MAG | Magnetometer |
| NASA | National Aeronautics and Space Administration |
| PSR | Permanently Shadowed Region |
| SEE | Secondary Electron Emission |
| SEP | Solar Energetic Particle |
| SIDE | Suprathermal Ion Detector Experiment |
| UV | Ultraviolet |
| VUV | Vacuum Ultraviolet |

# Abstract

As the Artemis program advances toward the lunar south polar region and permanently shadowed regions (PSRs), understanding the lunar electrical environment is increasingly important for assessing electrostatic hazards to astronauts, robotic systems, scientific instruments, and surface infrastructure. Lunar charging has traditionally been examined in terms of near-surface current balance driven by solar radiation, solar-wind and terrestrial magnetospheric plasmas, and secondary-electron emission. However, persistent shadow, cryogenic temperature, extremely low electrical conductivity, and prolonged charge-relaxation times in PSRs create conditions under which energetic-particle-induced charge may persist and accumulate beneath the lunar surface.

This Technical Memorandum reviews the environmental and physical context of lunar surface and subsurface charging while addressing a more fundamental unresolved question: how can electrostatic charge be generated, spatially separated, retained, and accumulated at depth within lunar regolith? Existing surface-charging models describe transitions between positive and negative surface potentials through current balance, and previous deep-dielectric-charging models predict that solar energetic particles and galactic cosmic rays may generate subsurface electric fields capable of approaching or exceeding dielectric-breakdown thresholds. However, the microscopic physical connection between an incident energetic particle and the volumetric charge source assumed in these macroscopic models has not yet been established.

Energetic-particle penetration and energy deposition determine where interactions and charge-carrier generation may occur, but they do not uniquely determine the charge that remains. Incident electrons, protons, and their secondary particles undergo stochastic elastic and inelastic interactions that may result in primary-particle stopping, implantation, backscattering, or transmission; secondary-electron generation and transport; electron recollection and reabsorption; prompt recombination; trapping; and escape across grain, pore, and bulk-regolith boundaries. Substantial energy may therefore be deposited without retention of the incident primary charge, while electron–hole separation and intergrain charge separation may remain even when the net charge of a larger control volume is unchanged.

To connect these microscopic processes to macroscopic subsurface charging, this memorandum defines the required microscopic input as the average signed retained-charge distribution per unit depth per incident particle, resolved by particle species and incident energy. Retained charge includes the charge of stopped or implanted primary particles, residual electron–hole separation, and intergrain charge separation remaining after primary- and secondary-particle transport, recollection, reabsorption, recombination, trapping, and escape. When combined with the corresponding incident particle flux and energy spectrum, this retained-charge response provides a physically motivated basis for constructing a depth-dependent volumetric charge-source rate. This source may then be coupled with charge continuity, conduction, dielectric relaxation, and Poisson's equation to describe the temporal evolution of subsurface charge density and internal electric fields.

The proposed formulation relies on three working assumptions: grain- and pore-scale heterogeneity can be represented by a depth-dependent ensemble average over a sufficiently large control volume; retained-charge responses may initially be superposed approximately linearly; and microscopic retained-charge formation occurs more rapidly than subsequent bulk conduction and dielectric relaxation. The spatial scale required for valid averaging, the charging level at which accumulated electrostatic potential modifies secondary-electron escape and redistribution, and the degree of timescale separation under cryogenic lunar conditions remain unresolved.

The principal quantities requiring further constraint are the charge contributions associated with primary-particle stopping, implantation, backscattering, and transmission, together with the fractions of generated electrons that recombine, are recollected, are reabsorbed by neighboring grains, become trapped, or escape from the bulk regolith. Their dependence on mineralogy, impact-glass content, grain and pore geometry, temperature, and pre-existing electrostatic potential must be determined through particle-resolved modeling and cryogenic high-vacuum irradiation experiments. Observations from Surveyor, Apollo, Lunar Prospector, Kaguya, and ARTEMIS are considered as environmental and observational constraints rather than direct measurements of subsurface retained charge.

By defining the missing microscopic-to-macroscopic charge-source connection, this memorandum provides a framework for investigating how subsurface charge distributions develop, whether they influence near-surface electric fields and surface charging behavior, and under what conditions they may produce dielectric breakdown and long-term regolith modification. Establishing this connection is necessary for evaluating electrostatic dust transport, surface-system contamination, astronaut exposure, and charge-mitigation requirements for sustained lunar polar exploration.

# Introduction

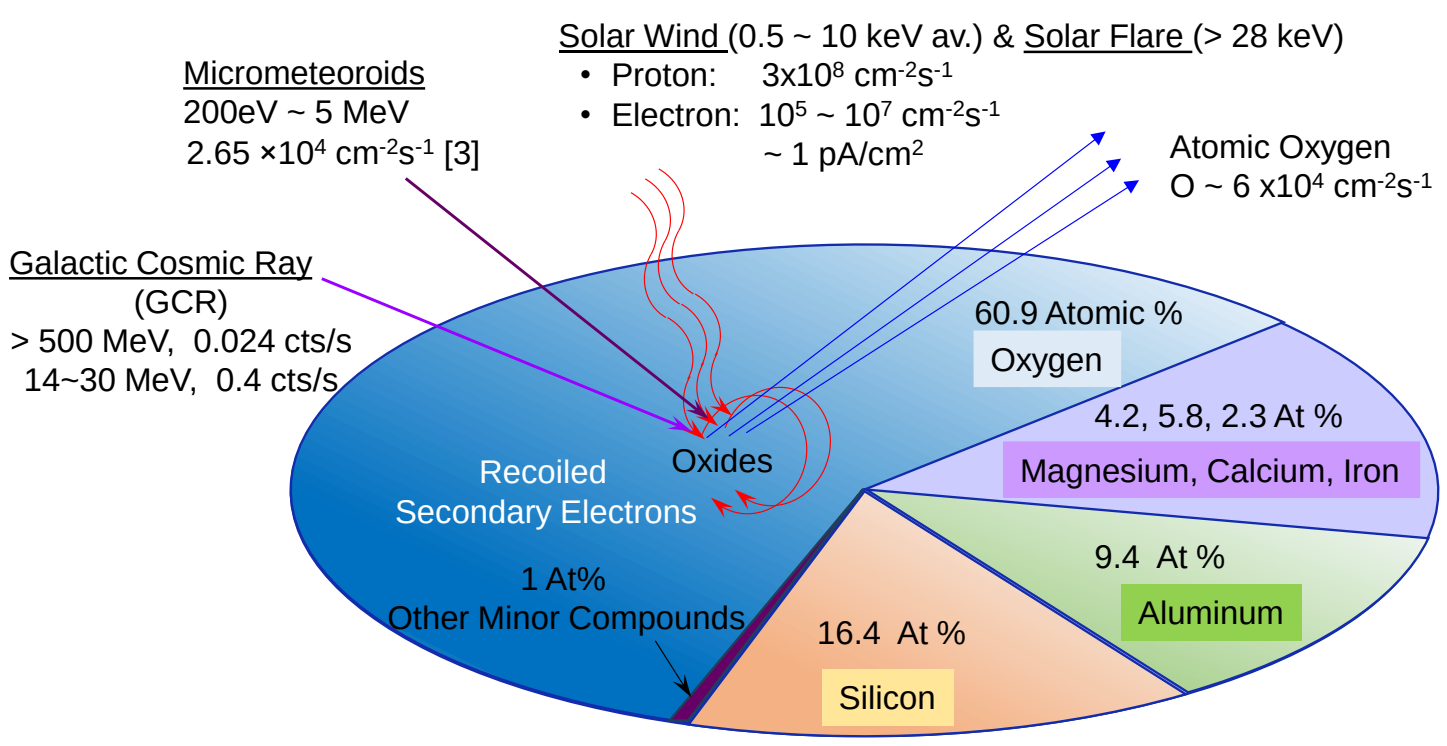

*Figure 1. Interaction patterns of lunar regolith with solar flares, solar wind, micrometeoroids, and galactic cosmic rays [1].*

The Moon lacks a substantial atmosphere, which leaves a high vacuum of about $10^{-11}$ torr, and it has no global intrinsic magnetic field, so its surface is directly exposed to solar radiation, solar wind, solar flares, terrestrial magnetospheric plasma, energetic particles, and micrometeoroid impacts. Under these conditions lunar regolith continuously exchanges charged particles with the surrounding space environment and becomes electrostatically charged. Because the lunar surface is in a vacuum environment, collisional charge redistribution and neutralization are extremely limited, and the low electrical conductivity of the regolith allows accumulated charge to persist for extended periods. Figure 1 illustrates representative interaction pathways between lunar regolith and solar wind, solar flares, galactic cosmic rays, and micrometeoroids. It shows that electrons and ions can be generated, emitted, implanted, or retained at the regolith surface and within near-surface grains, providing the physical basis for both surface charging and subsurface charge accumulation [1].

Lunar charging has traditionally been examined in terms of surface charging governed by the current balance among photoelectron emission, ambient electron and ion impact, and secondary electron emission. Additional processes, including energetic-particle-induced charge trapping and retention, micrometeoroid impact, and triboelectric charging, further modify the charging behavior and eventually the charge density that generates local potential differences.

The lunar charge population is created dominantly by the impacts of energetic electrons, ions, and high-energy photons from the solar wind and solar flares, from micrometeoroid impacts, and from galactic cosmic rays [1]. In the resulting charge-density balance, secondary and tertiary electron emission from regolith particles takes place and sequentially modifies the electrostatic field, so that the interaction allows not only surface charge but also charge penetration through the regolith particles as a secondary modifying process.

When energetic particles and high-energy photons strike the regolith, the impact energy greatly exceeds the binding energies of the target atoms and liberates not only the electrons of the outermost shell of the atomic structure but also the electrons of the inner shells. When the energies of the incident particles and photons are excessively high, the impact processes induce Bremsstrahlung, mainly as X-ray emission, and produce energetic secondary electrons by Compton scattering. These induced X-rays and Compton electrons still carry a great deal of energy and therefore have secondary and tertiary effects, and it is these secondary and tertiary effects that set the causal chain of charge development in depth; the generation and escape of the resulting secondary populations are treated in Section 2.3.

The sign and magnitude of the resulting surface potential depend on the plasma environment as much as on the illumination, as set out in Chapter 1. Local factors are superimposed on this large-scale dependence. The composition, grain size, temperature, conductivity, and surface roughness of the regolith all modify electron emission and charge retention through photocoupling effects such as the photoelectric effect and Compton scattering, and their individual roles are examined in Section 1.6.

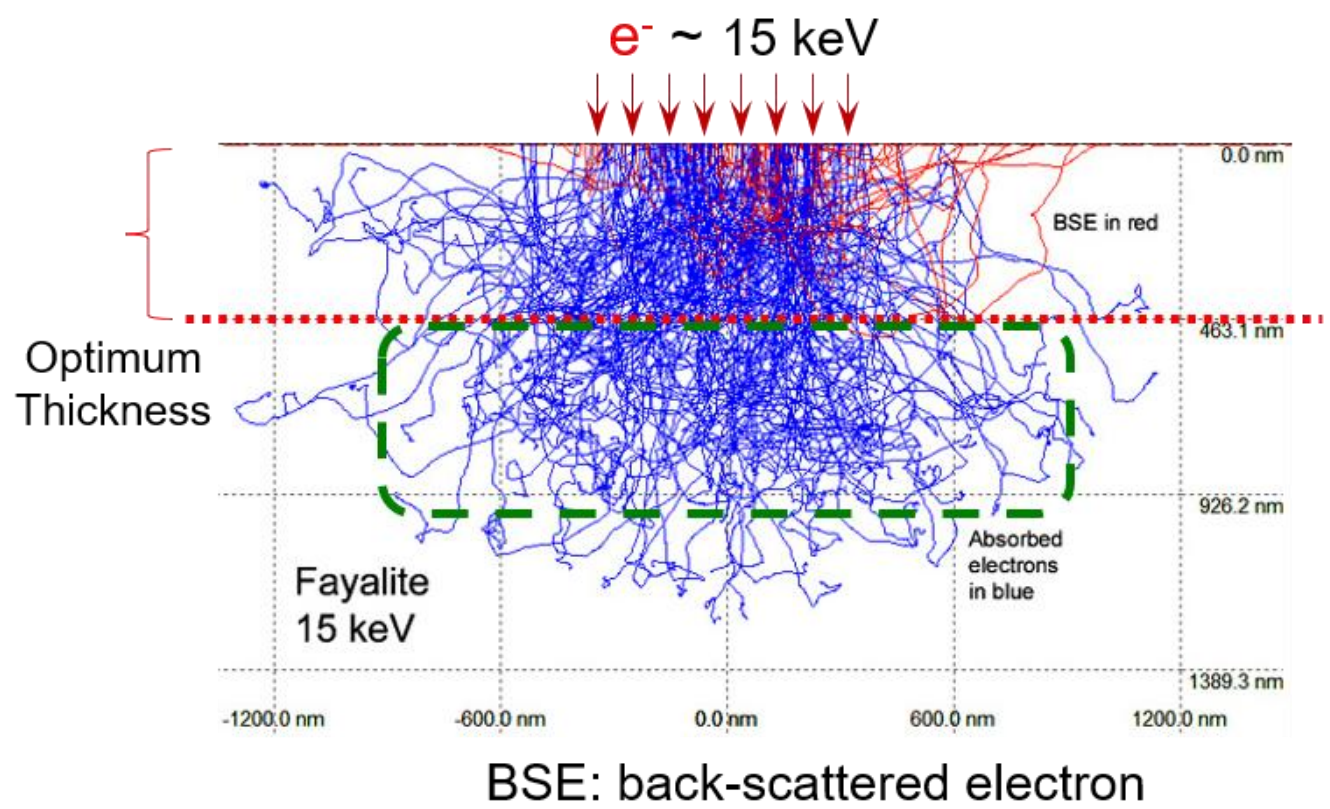

*Figure 2. Monte Carlo analysis of the electron transition from bound to free states under the bombardment of 15 keV electrons onto fayalite ($Fe_2SiO_4$). Backscattered electrons are shown in red and forward-scattered, absorbed electrons in blue.*

The energies involved set the depth over which these processes act. The electrons and protons of the solar wind are of the order of 1 keV [2, 3], whereas the electrons of a solar flare carry more than 100 keV and its protons range from about 15 MeV to several GeV. These levels of solar-flare energy by themselves have a formidable impact in yielding free electrons from the regolith. Figure 2 shows the result of a Monte Carlo analysis for 15 keV electrons impinging on fayalite ($Fe_2SiO_4$). Apart from the backscattered electrons, shown in red and confined to the near-surface layer, the majority of the impact energy is used to yield forward-scattered electrons, shown in blue, whose trajectories reach a depth of approximately 1.4 μm. If this result is extended to the case of solar-flare electrons at 100 keV, the penetration depth of the liberated electrons may reach the tens-of-micrometer range for electrons of that energy. The energetic protons of a solar flare show a similar nature of interaction, but they carry energies at the GeV level, which is sufficient to crush the nucleus of an atom, and their penetration is correspondingly deeper; the characteristic depth scales for both species are quantified in Section 2.5. When the nucleus is disrupted in this way, the induced emission of X-rays and gamma rays by Bremsstrahlung is a prevalent feature of the post-interaction, on top of the exceptionally high yield of free electrons released directly by the primary interaction and by the secondary interactions of the X-rays, gamma rays, and energetic electrons.

These considerations indicate that lunar charging is governed by the time-varying solar and radiation environments. Although the current-balance framework explains most observed variations in lunar surface potential, it does not by itself describe how charge is generated, separated, and retained within the subsurface. Under irradiation by high-energy photons and energetic particles, subsurface charging can arise through energy deposition, ionization, and secondary-particle generation within the regolith.

The cascade described above does not deposit its charge at a single surface: each generation of carriers is created deeper than the last, so that charge is redistributed over a finite depth range rather than confined to the outermost layer. GCRs are relatively sparse, but their high energies allow them to produce extensive ionization cascades. Although most generated carriers recombine, are reabsorbed, or escape, a small residual fraction may remain trapped or spatially separated in the highly insulating regolith. Repeated over geological timescales, this residual charge may accumulate and contribute to deep dielectric charging.

The impact effect by micrometeoroids is twofold. One is the direct ionization by the impact energy that releases electrons. The other is the covering effect on the existing surface charge by the pulverized regolith thrown over it. If these activities are counted over a long period of time, the development of charge potential at depth is by no means negligible. The deposited charges within the highly insulating regolith can hardly escape but are kept in place for a long period, so that they become trapped and accumulate within the subsurface. The internal electric fields that such a charge distribution generates, and the conditions under which they may lead to dielectric breakdown, are developed in Chapter 2. Understanding these subsurface charging processes has therefore become increasingly important, since it can offer some level of guidance for the design of systems and probes for the Artemis III mission activities.

This paper reviews the principal physical processes governing both lunar surface and subsurface charging. In addition, this paper discusses energetic particle penetration, charge accumulation, and the potential for internal electric-field development and dielectric breakdown. Observational results from Surveyor, Apollo, Lunar Prospector, Kaguya, and ARTEMIS (Acceleration, Reconnection, Turbulence and Electrodynamics of the Moon's Interaction with the Sun) are compared to provide an interpretation of lunar surface charging across diverse lunar environments.

# 1 Lunar Charging Environment

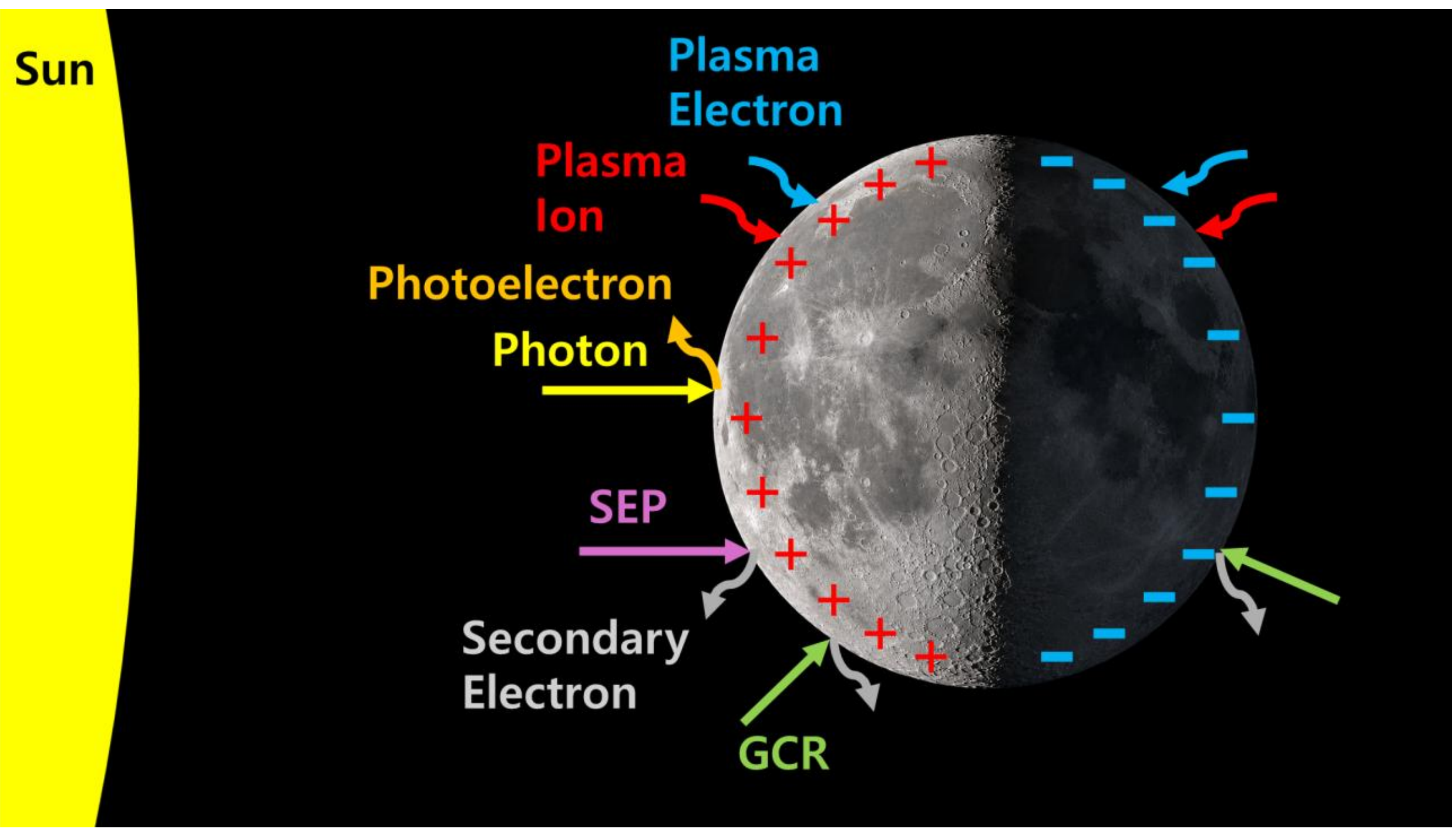

*Figure 3. Lunar surface charging overview.*

The electrostatic state of the lunar surface is determined by the competition among several charge-exchange processes acting simultaneously. Solar illumination, ambient plasma electrons and ions, and the fluxes and energies of energetic particles collectively control the surface potential. Figure 3 provides an overview of these major lunar surface-charging mechanisms. The relative strengths of these processes vary with the surrounding environment. This chapter introduces the principal lunar charging environments and their associated mechanisms. Section 1.1 presents the current-balance framework that connects the individual processes. Sections 1.2–1.4 describe the photon, energetic-particle, and plasma environments, together with the emission and collection mechanisms they induce. Section 1.5 discusses triboelectric charge transfer, Section 1.6 addresses the relevant properties of lunar regolith, and Section 1.7 summarizes temporal variations in the overall charging environment.

## 1.1 Current Balance Equation

The electric potential of the lunar surface is governed by the net current density flowing to and from the surface. At the macroscopic scale, this current can be approximated as the sum of four principal contributions:

$$J_{total}(\Phi) \approx J_{ph}(\Phi) + J_i(\Phi) + J_e(\Phi) + J_{see}(\Phi)$$

Where $\Phi$ is the surface potential relative to the ambient plasma. Here, $J_{\mathrm{ph}}$ denotes the photoelectron-emission current, $J_{\mathrm{i}}$ and $J_{\mathrm{e}}$ denote the ion and electron currents, respectively, and $J_{see}$ represents the current associated with secondary-electron emission. Each contribution depends on $\Phi$, because the surface potential modifies the collection or escape of charged particles by electrostatically attracting or repelling them. The quasi-steady surface potential $\Phi_{\mathrm{s}}$ is therefore determined by the current-balance condition

$$J_{\mathrm{tot}}(\Phi_{\mathrm{s}}) = 0.$$

Early treatments of lunar charging approximated the Moon as a globally smooth, absorbing sphere or as an infinite planar surface immersed in a homogeneous plasma [4, 5]. Under these assumptions, simple analytic expressions can be combined with a prescribed photoelectron current to yield dayside and nightside potentials as functions of solar-zenith angle and plasma parameters [4, 6]. Such models reproduce the broad trend of a mildly positive dayside (of order +10 V) and a more negative nightside (tens of volts negative in the solar wind, larger in the magnetotail) [6, 7]. However, they neglect the strong small-scale variability introduced by rough topography, compositional heterogeneity, and discrete dust grains [8, 9].

To capture this variability, more recent work has adopted patch and grain models [8, 10, 11]. In patch models, the surface is divided into facets with different illumination conditions, material properties, and plasma exposure. Each patch $p$ satisfies its own current-balance condition:

$$\sum_k J_k^{(p)}\left(\Phi^{(p)}\right) = 0.$$

Here $J_k^{(p)}$ denotes the contribution from mechanism $k$ on patch $p$, and $\Phi^{(p)}$ is the local potential on that patch. Neighboring patches are electrostatically coupled through overlapping sheath fields, so contrasts in illumination, composition, or plasma conditions generate lateral electric fields [8, 11]. Where patch sizes approach the local Debye length, these lateral fields produce intense small-scale potential structures across shadow boundaries, near topographic edges, and above crustal magnetic anomalies [8, 9, 11]. The following sections discuss the environmental factors governing lunar surface charging and the physical relationships through which they affect the surface potential.

## 1.2 Photon Environment

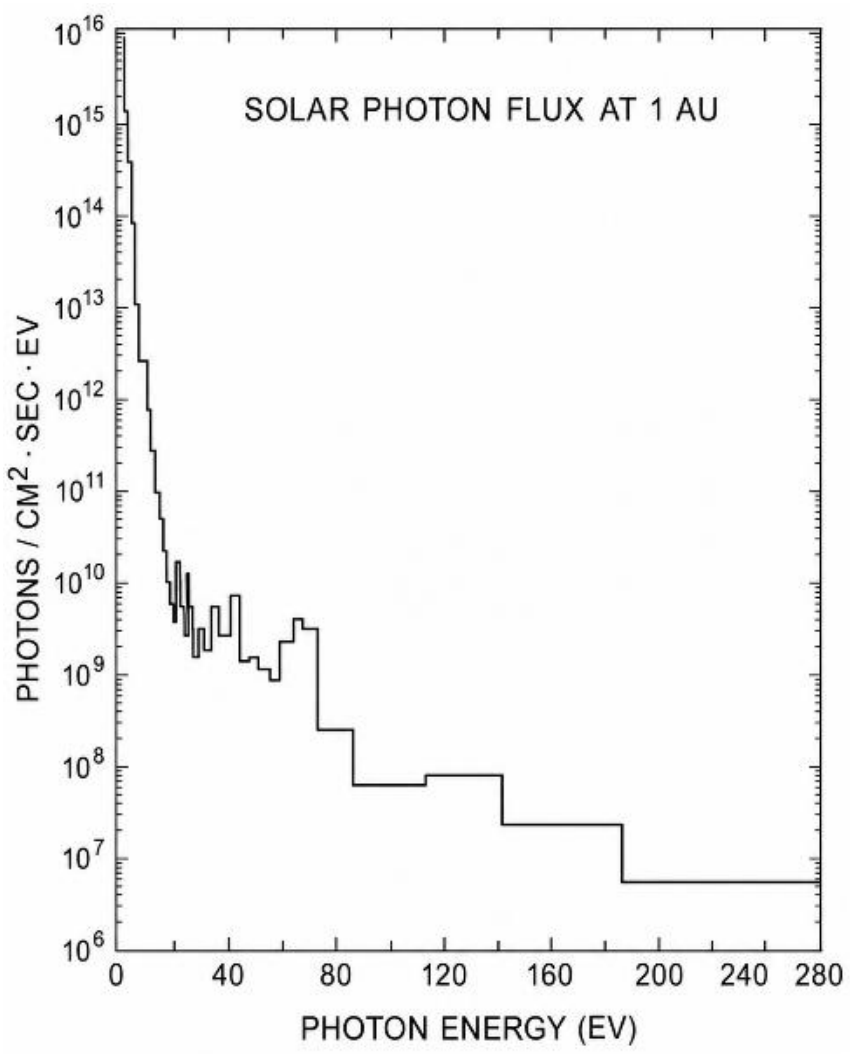

*Figure 4. Measured solar photon flux as a function of photon energy at 1 AU [12].*

Solar radiation consists of photons spanning a broad range of wavelengths and energies. Figure 4 shows the measured solar photon flux at 1 AU as a function of photon energy, indicating that the photon flux varies strongly with photon energy. On Earth, photons in the VUV (Vacuum Ultraviolet) range are absorbed by the atmosphere. However, this atmospheric absorption does not occur on the Moon because it lacks an atmosphere. Therefore, when estimating the photon environment at the lunar surface, atmosphere-free measured values of the solar spectrum must be used [12, 13].

### 1.2.1 Photoemission

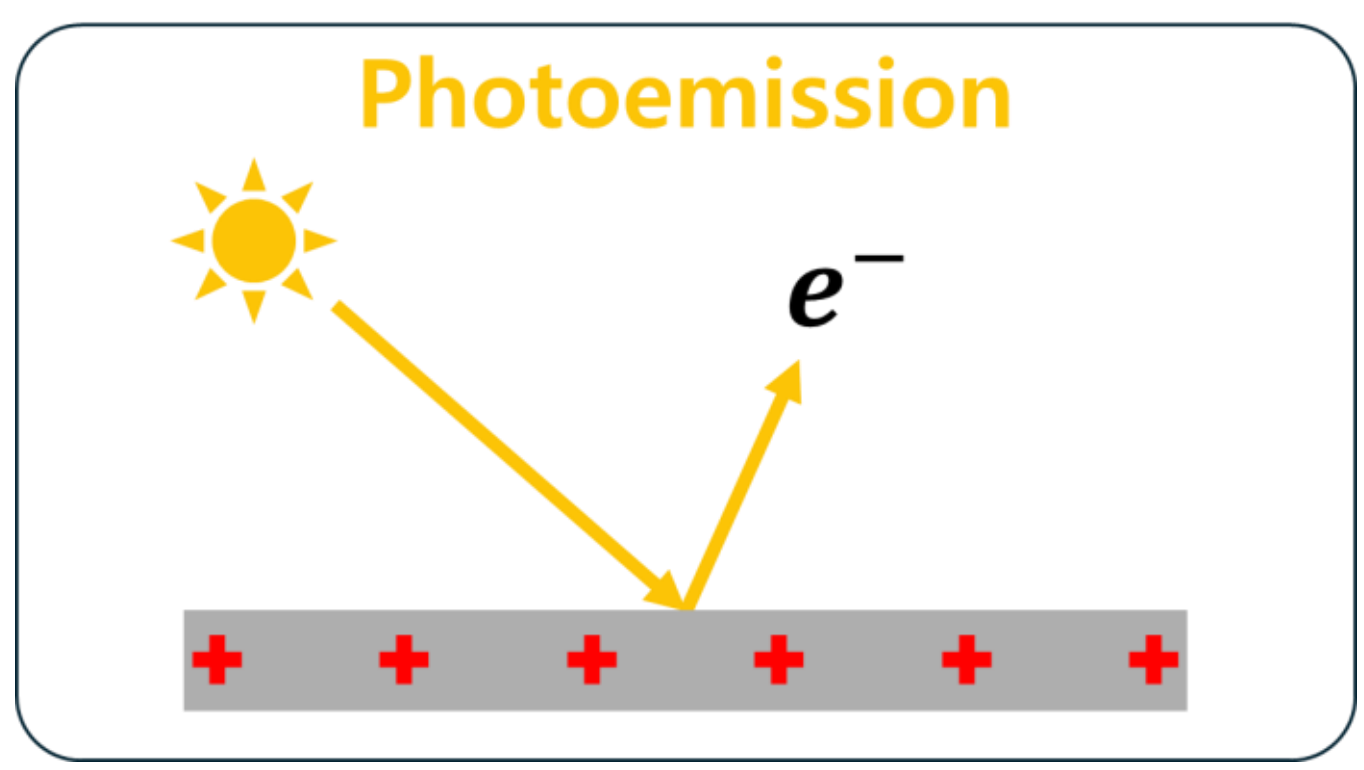

*Figure 5. Conceptual diagram of photoelectron emission from the lunar surface.*

Photoemission occurs when an incident photon is absorbed by the material and transfers sufficient energy to an electron in an occupied valence, defect, or surface state for that electron to overcome the effective surface barrier and escape into vacuum. The required photon energy is therefore set by the material-dependent photoemission threshold W, which is related to the electronic structure and surface condition of the material. The emitted electrons are referred to as photoelectrons. Because their departure removes negative charge from the surface, photoemission drives an illuminated surface toward positive charging, as shown schematically in Figure 5. Under the extremely tenuous lunar atmosphere, photoelectrons can travel over appreciable distances without collisions with neutral gas molecules; depending on the local electric field, they escape, are recollected by the emitting surface, or are deposited on nearby grains or structures [8, 14]. Photoemission can therefore produce both local surface charging and charge redistribution within the regolith [8, 14].

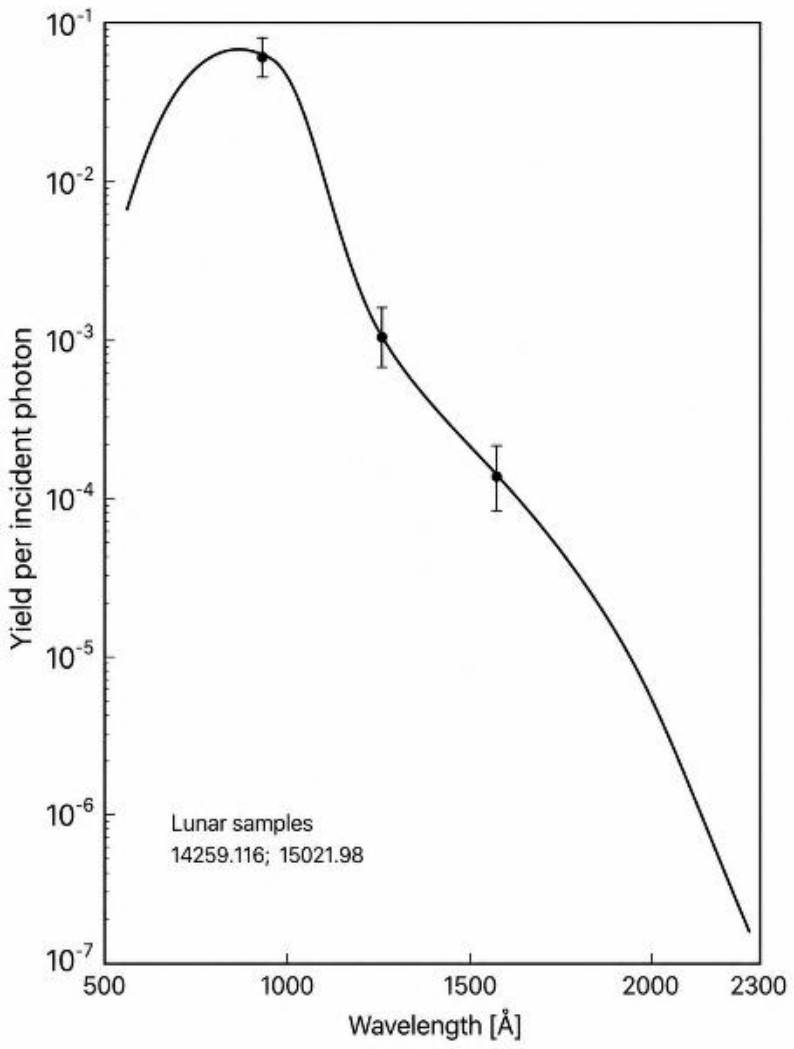

*Figure 6. Differential photoelectric yield of Apollo lunar dust [15].*

For a photon of wavelength λ, the photoelectron yield $Y_{\mathrm{ph}}(\lambda)$ is defined as the number of electrons emitted per incident photon. This yield is wavelength dependent because photoemission involves several successive processes: photon absorption, excitation of an electron, transport of the excited electron toward the surface, and escape across the surface barrier. Even when the photon energy exceeds the threshold, the yield generally remains below unity because some photons are reflected, while others are absorbed too deeply for the excited electrons to reach the surface before losing energy, recombining, or becoming trapped. The yield also depends on material composition, grain structure, roughness, defects, contamination, and electrostatic condition, and must therefore be determined experimentally. Figure 6 reproduces the differential yield measured for Apollo 14 and 15 dust samples over 500–2300 Å. The yield rises steeply with decreasing wavelength, reaches a maximum of about 9% near 900 Å, corresponding to 14 eV, and falls to roughly 1% at 584 Å, or 21 eV [15]. The photoemission threshold is approximately 5 eV, corresponding to 2480 Å [15, 16]. Analytical sheath models commonly adopt a slightly higher effective threshold of W ≈ 6 eV [12].

For a given incident photon-flux spectrum $F_{\gamma}(\lambda)$ the zero-potential photoelectron current density is obtained by weighting the photon flux by the wavelength-dependent yield over the relevant UV/EUV band:

$$J_{\mathrm{ph},0} = e\int_{\lambda_{\min}}^{\lambda_c} Y_{\mathrm{ph}}\,(\lambda)\,F_\gamma(\lambda)\,d\lambda,$$

where $e$ is the elementary charge and the upper integration limit is determined by the threshold wavelength,

$$\lambda_c = \frac{hc}{W}$$

Only photons satisfying $\lambda < \lambda_c$, or equivalently $h\nu > W$, can contribute directly to photoemission. Although the solar photon flux generally decreases toward shorter wavelengths, the photoelectron yield varies strongly with photon energy; consequently, the photoelectron current must be evaluated by integrating the product of the photon flux and yield rather than from either quantity alone.

The photon flux that drives photoemission varies systematically over the solar cycle, as discussed in Section 1.7.3. When the measured yield spectrum is integrated over the solar VUV–EUV spectrum, the resulting zero-potential photoelectron current density is 4.5 $\mu A\ m^{-2}$, equivalent to an emission rate of about $3 \times 10^9$ electrons $cm^{-2}\ s^{-1}$, near local noon at solar minimum; it rises to approximately 15 $\mu A\ m^{-2}$ near solar maximum and reaches about 40 $\mu A\ m^{-2}$ during strong solar flares [10, 15]. These values provide a characteristic reference scale for the outward electron current in lunar surface current-balance models.

Laboratory measurements further indicate that lunar photoelectrons have a characteristic mean kinetic energy of approximately 2 eV, consistent with the effective photoelectron temperatures $T_{ph} \approx 1$–3 eV commonly used in analytical sheath models [10, 15]. Together with the corresponding emission current, these energies imply a photoelectron density immediately above a sunlit lunar surface of order $10^2\ cm^{-3}$. The emitted electrons form a near-surface photoelectron sheath with a characteristic Debye length of approximately 1 m and an electric field of order several volts per meter [10, 15].

When the surface potential is nonzero, the emitted photoelectron current is additionally controlled by the electrostatic barrier above the surface. A negative surface potential accelerates emitted electrons away from the surface, whereas a positive potential attracts them back and reduces the fraction that can escape. For a Maxwellian photoelectron energy distribution with temperature Tph, the escaping photoelectron current can be approximated as [6]

$$J_{\mathrm{ph}}(\Phi) = \begin{cases} J_{\mathrm{ph0}}, & \Phi < 0 \\ J_{\mathrm{ph0}} \exp\left(-\dfrac{e\Phi}{k_B T_{\mathrm{ph}}}\right), & \Phi > 0 \end{cases}$$

## 1.3 Energetic Particle Environment (SEP, GCR)

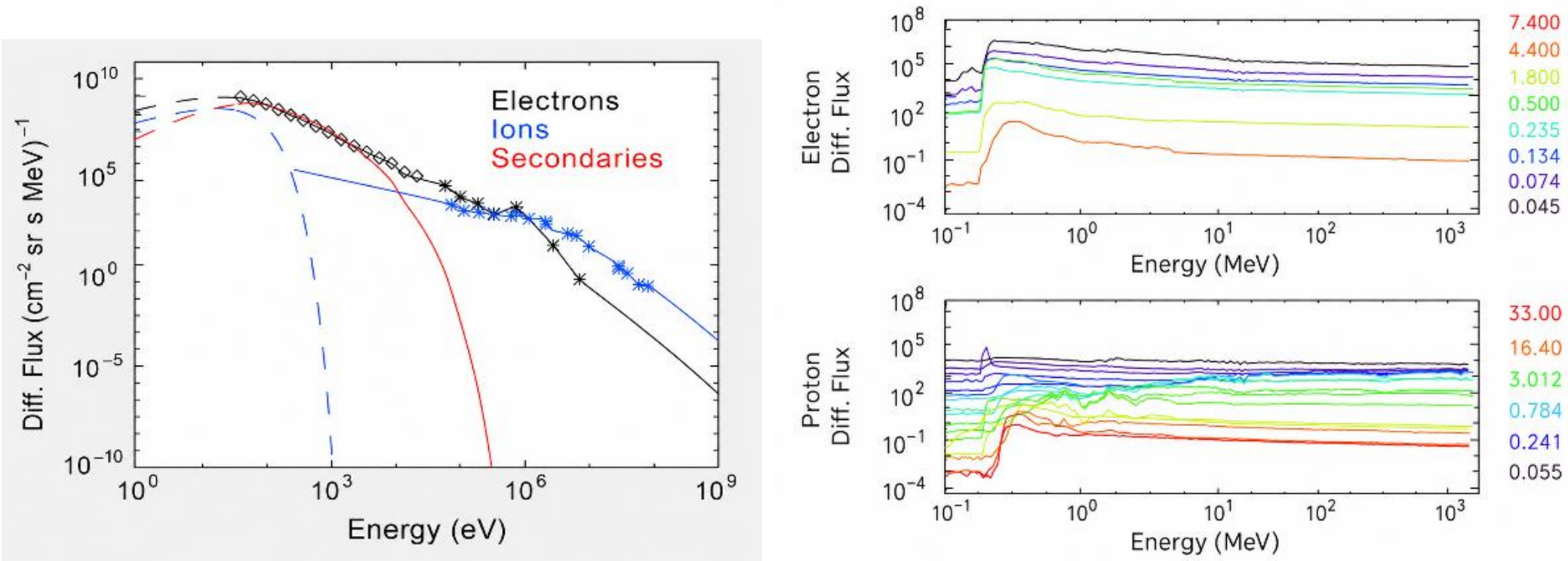

*Figure 7. (a) SEP's energy and flux ranges of different particle species [17]. (b) Differential flux with electron and proton from SEP [17].*

During solar energetic particle (SEP) events, the Moon is exposed to intense fluxes of energetic electrons and ions, spanning roughly tens of keV to several MeV for electrons and tens of keV to tens of MeV for protons [17]. As shown in Figure 7, the differential flux falls steeply with increasing energy, so the incident flux is dominated by the low-energy end of the spectrum. In the largest events the electron differential flux peaks near 0.2 MeV at values of order $10^5$ (cm$^2$ s sr MeV)$^{-1}$, while the proton differential flux near a few MeV reaches values of order $10^3$ (cm$^2$ s sr MeV)$^{-1}$ [17].

Galactic cosmic rays (GCRs) are high-energy charged particles, including atomic nuclei and electrons, that originate primarily from astrophysical sources such as supernova-driven shocks. Near Earth's orbit, GCRs consist of approximately 83% protons, 13% alpha particles, 1% nuclei heavier than helium, and 3% electrons [18]. The GCR proton spectrum typically peaks near an energy of approximately 200 MeV [19]. GCR flux varies over the solar cycle because the heliospheric magnetic field modulates the penetration of cosmic rays into the inner solar system. Consequently, stronger solar magnetic activity provides greater shielding and reduces the GCR flux, whereas weaker activity allows a larger GCR flux to reach the Moon.

### 1.3.1 Secondary Electron Emission

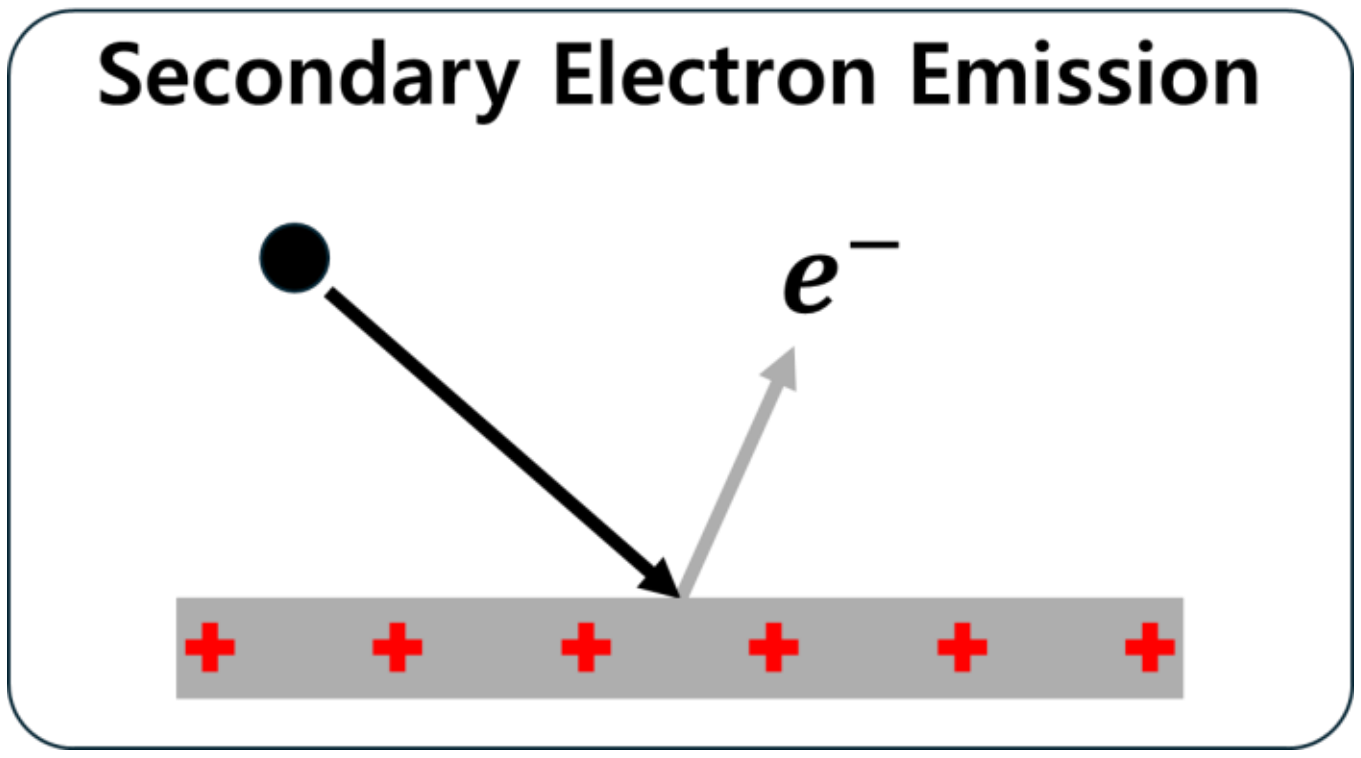

*Figure 8. Conceptual diagram of secondary electron emission from the lunar surface.*

When an energetic particle strikes a solid, it transfers part of its kinetic energy to the electrons of the target. These interactions excite or ionize bound and conduction electrons, producing low-energy electrons within the material. Electrons generated close enough to the surface migrate toward it and escape if they retain enough energy to overcome the surface potential barrier. These emitted target electrons are referred to as secondary electrons. Figure 8 shows this process.
When a charged particle penetrates the material, its trajectory may be redirected toward the surface through large-angle scattering within the solid. A particle that subsequently leaves the surface after such deflection is referred to as a backscattered electron or backscattered ion, depending on the incident species. Within the material, charged particles undergoing backscattering and energetic secondary electrons may cause additional collisions and generate further secondary electrons, thereby contributing to surface charging.

## 1.4 Plasma Environment

Under nominal solar-wind conditions near 1 AU, outside of the Earth's magnetotail, the median electron and proton temperatures measured over a decade of Wind observations are approximately $T_e \sim 11.9$eVand $T_i \sim 4.3$eV. The bulk flow speed ranges from roughly 300 km $s^{-1}$ in slow wind to above 700 km $s^{-1}$ in fast streams, with the nominal value taken here as $V_{sw} \sim 500\ \mathrm{km\,s^{-1}}$ [2, 3]. The flow is both supersonic and super-Alfvénic in most cases. As a result, disturbances cannot propagate upstream, and the solar wind interacts directly with obstacles such as the lunar surface, leading to the formation of plasma sheaths and downstream wake structures.
Downstream of the Moon the solar wind is absorbed by the lunar surface rather than deflected around it, leaving a plasma void known as the lunar wake. Plasma densities in the central wake fall by more than an order of magnitude below the ambient solar-wind value [20]. The cavity refills from its flanks, and because electrons are far more mobile than ions they enter the void first; the resulting charge separation establishes an ambipolar electric field that retards the electrons and accelerates ions inward. This ambipolar potential drop amounts to several hundred volts across the wake boundary, and the central wake is observed at roughly −300 V relative to the surrounding solar wind, with strongly reduced electron density and elevated electron temperature [21]. This depleted, negatively biased environment is the principal reason for the strongly negative charging of the lunar nightside in the solar wind.
When the Moon passes through Earth's magnetotail, the plasma incident on the lunar surface is no longer supplied directly by the solar wind but is instead replaced by plasma trapped within the terrestrial magnetosphere. The magnetotail is broadly divided into the tail lobes, plasma sheet boundary layer, and central plasma sheet, and the effects of these regions on lunar surface charging differ substantially because of their distinct plasma densities, temperatures, and particle-energy characteristics. The Moon encounters different magnetotail plasma regimes along its orbit, ranging from the tail lobes to the central plasma sheet; the geometry of this passage is described in Section 1.7.2.
The magnetotail lobes correspond to the northern and southern high-latitude regions of the magnetotail, where Earth's magnetic field partially shields the Moon from direct solar-wind plasma access. These regions are characterized by extremely tenuous plasma, with typical densities below 0.1 $cm^{-3}$, and by a strong and relatively steady magnetic field [22]. Because the ambient electron and ion currents are greatly reduced under these low-density conditions, weaker current sources can become comparatively important.
In contrast, the central plasma sheet contains hotter and denser magnetospheric plasma than the lobes. Typical plasma sheet number densities are on the order of 0.1–1 $cm^{-3}$, and average ion temperatures are generally a few keV, substantially higher than the corresponding electron temperatures [23]. The elevated electron density and temperature in this region increase the ambient electron flux incident on the lunar surface. Negative surface potentials have been reported in this environment even on the sunlit dayside, in contrast to the positive values obtained under nominal solar-wind conditions; these observations are presented in Chapter 3.

Between the lobes and the central plasma sheet lies the plasma sheet boundary layer, a transitional region where lobe-like and plasma-sheet-like populations coexist at densities of order 0.1–2 $cm^{-3}$ [22]. This region is frequently associated with field-aligned beams and energy-dispersed particles generated by magnetotail current-sheet acceleration and magnetic reconnection processes. Observations show that field-aligned ion beams in the plasma sheet boundary layer have energies below ~40 keV during quiet Type-I beam events and can reach up to ~200 keV during more active Type-II beam events [24]. Therefore, when the Moon is located near the plasma sheet boundary layer, the particle flux incident on the lunar surface can change rapidly, leading to a more complex and time-dependent charging environment.

### 1.4.1 Electron Impact

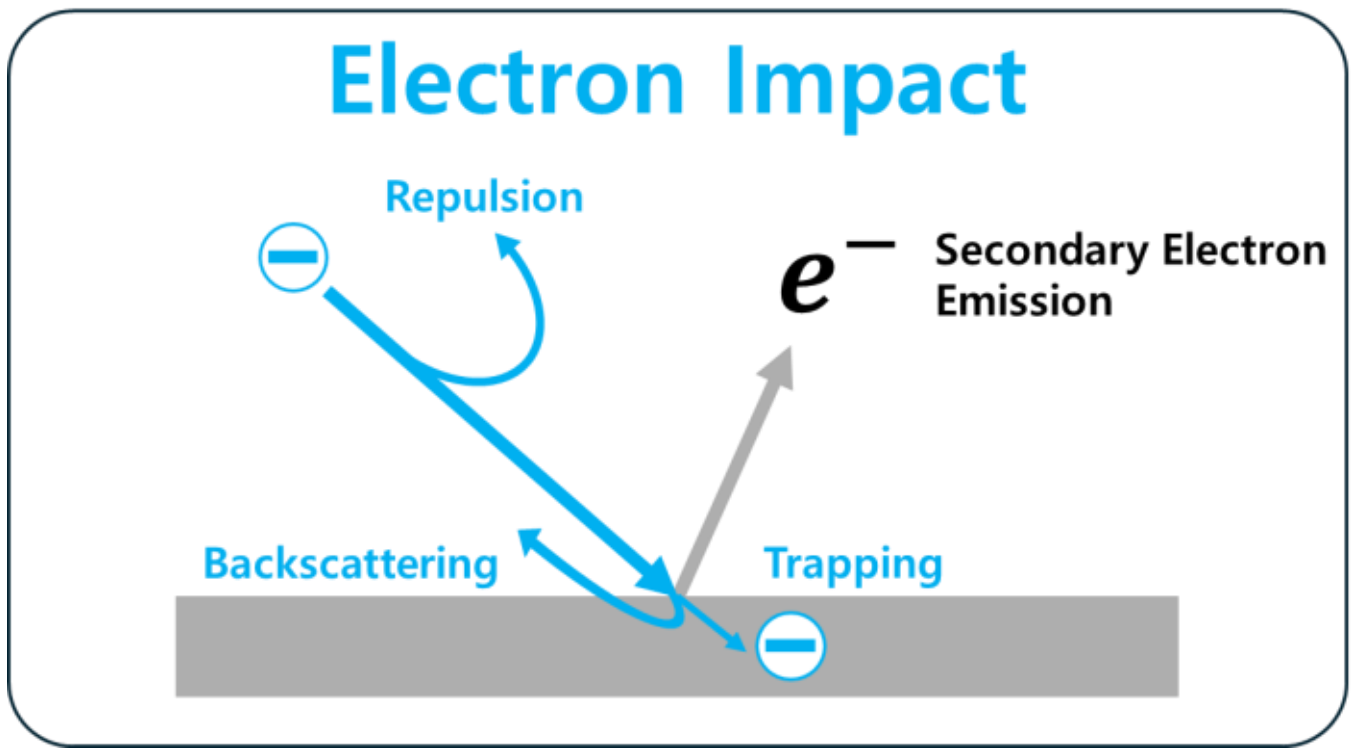

*Figure 9. Conceptual diagram of electron interactions with the lunar surface.*

As shown in Figure 9, electrons reaching the lunar surface can undergo several different interactions. Electron backscattering and secondary-electron emission have already been discussed in Section 1.3.1. Electrons that penetrate the material and lose kinetic energy through collisions become trapped in localized defect states such as dangling bonds, adding negative charge to the near-surface region.

For macroscopic surface-charging analysis, the key quantity is the balance between the electron current incident on the surface and that leaving it. In practice, however, it is difficult to quantify separately every process involved in particle interactions within the material. The relative contributions of these processes depend on the energy and species of the incident particles, as well as on the composition, roughness, contamination, and electrostatic condition of the surface. The net electron-emission response of lunar regolith is therefore generally characterized experimentally rather than by resolving each emission and scattering process individually. When electrons are incident on the surface, the total electron yield, defined as the ratio of the total outward electron-current density to the incident electron-current density, serves as the principal parameter for evaluating the resulting charging tendency:

$$\sigma(E) = \frac{J_{\text{out}}}{J_{\text{in}}}$$

where $J_{in}$ is the incident current density, $J_{\text{out}}$ is the emitted current density, and E is the primary electron energy. Although these outgoing populations arise through different microscopic mechanisms, they all remove negative charge from the surface and can therefore be combined for macroscopic charge accounting.

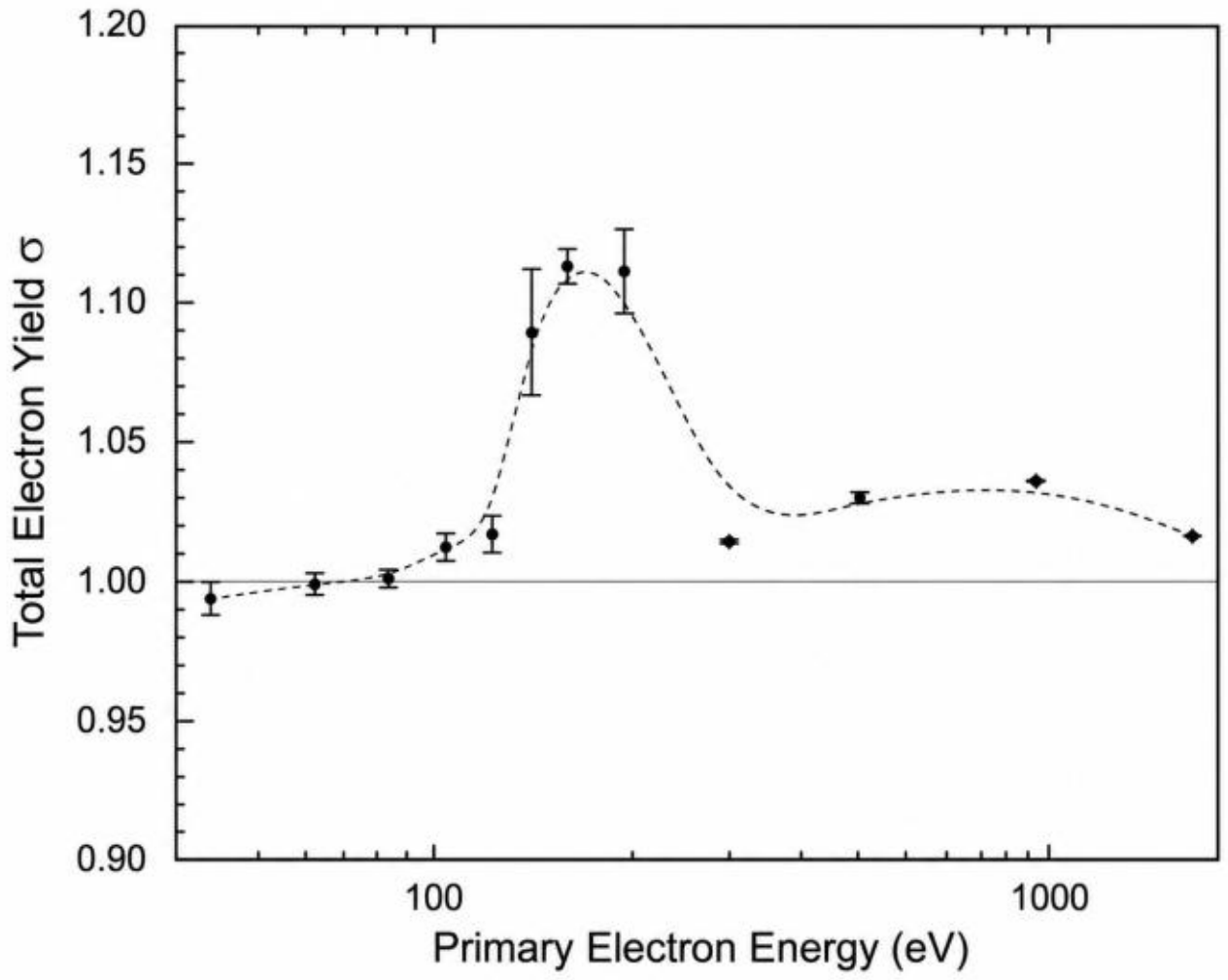

*Figure 10. Total electron yield as a function of primary electron energy for Apollo soil [25].*

The total electron yield depends strongly on the primary-electron energy and indicates the direction of electron-induced charging. When σ(E) < 1, fewer electrons leave the surface than arrive, resulting in net electron deposition and a negative charging tendency. When σ(E) > 1, the outward electron current exceeds the incident current, causing the surface to lose negative charge and producing a positive charging tendency. As the primary energy increases, the yield generally rises through a first crossover energy, reaches a maximum, and decreases again at higher energies. Figure 10 presents the experimentally measured total electron yield for Apollo-returned lunar soil. For lunar soil, the yield first crosses unity at approximately 80 eV, reaches a maximum of about 1.13, and then decreases with increasing incident electron energy [25]. Earlier measurements on other Apollo soils gave higher maxima of 1.4–1.6, a difference attributed to their larger glass content [15, 25].

Whether an electron reaches the surface in the first place is governed by the instantaneous surface potential: a negatively charged surface repels approaching electrons and suppresses the collected flux, whereas a positively charged surface accelerates them inward and enhances it. For a Maxwellian electron population of temperature $T_e$ and zero-potential current density $J_{e0}$, the collected electron current density as a function of surface potential is [6]:

$$J_e(\Phi) = \begin{cases} J_{e0} \exp\left(\frac{e\Phi}{k_B T_e}\right), & \Phi < 0 \\ J_{e0} \left[1 + \frac{e\Phi}{k_B T_e}\right], & \Phi > 0 \end{cases}$$

### 1.4.2 Ion Impact

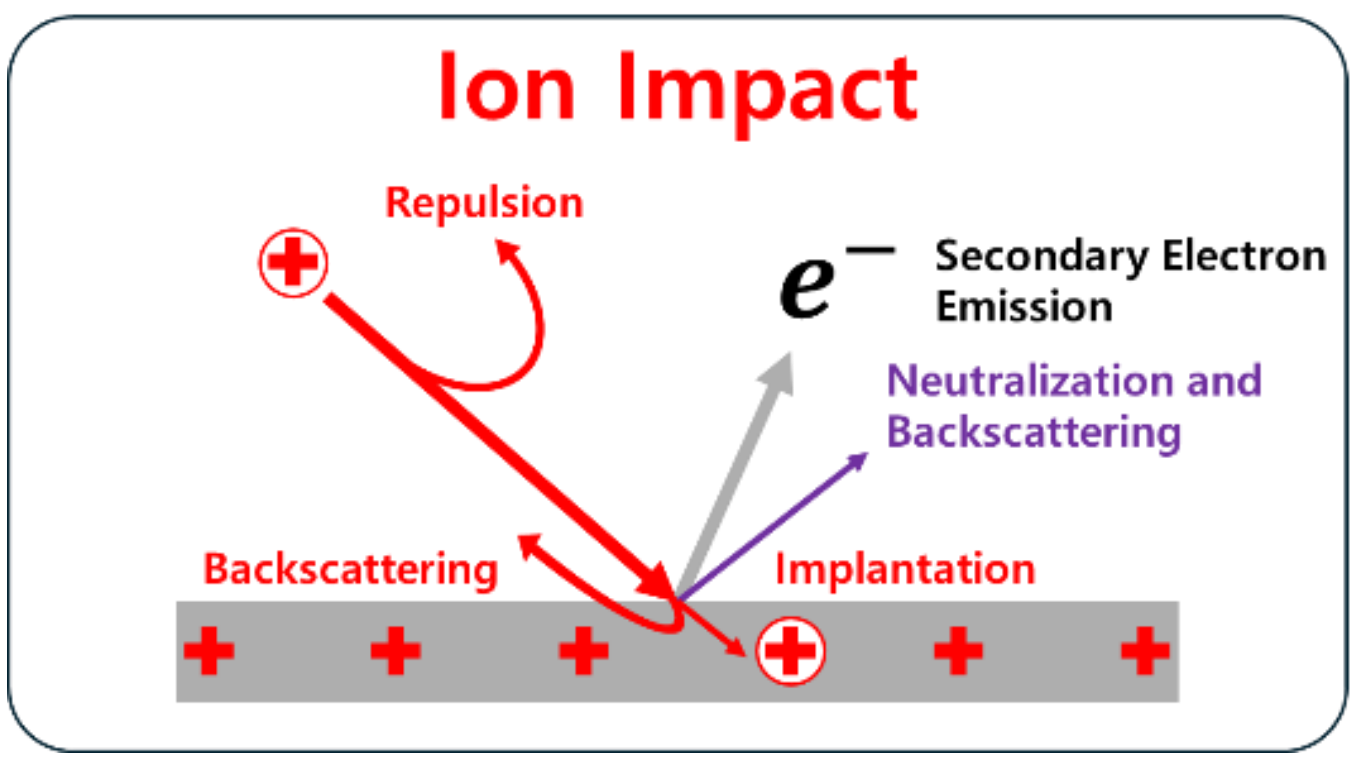

*Figure 11. Conceptual diagram of ion interactions with the lunar surface.*

As shown in Figure 11, ions reaching the lunar surface can undergo processes analogous to those produced by electron impact, including secondary-electron emission and backscattering. For an incident positive ion, charge exchange with electrons in the solid neutralizes the ion. If the resulting neutral atom is then backscattered and escapes from the surface, it carries away an electron acquired from the material, leaving an electron deficit and shifting the surface potential in the positive direction. If the positive ion instead loses sufficient energy and becomes implanted in the near-surface material, its retained charge also contributes to positive charging. Observations constrain how the incident flux divides between these outcomes: the measured reflection efficiency for solar-wind protons returning as ions is of order 1%, whereas the flux returning as neutralized hydrogen after charge exchange is substantially higher [26]; the remainder is implanted within the uppermost ≈100 nm of the grains, where the lattice is defect-rich [27].
Because most ions incident on the lunar surface are solar-wind $H^+$, Figure 11 shows a positive ion. Negative ions are not a significant component of the incident flux; hydroxyl at the lunar surface is instead formed in situ when implanted solar-wind protons react with metal oxides in the regolith [28]. Nevertheless, if negative ions were to reach the surface, the corresponding processes would act in the opposite sense: neutralization followed by scattering would leave an additional electron in the surface and shift the potential in the negative direction, while implantation of a negative ion would similarly deposit negative charge in the material.
For macroscopic surface-charging analysis, it is difficult to quantify separately the contributions of secondary emission, charge exchange, neutralization, backscattering, and implantation. As with electron impact, the net charging response of lunar regolith under ion impact is therefore generally characterized experimentally rather than by resolving each microscopic interaction individually.
Whether an ion reaches the surface in the first place is governed by the instantaneous surface potential: a positively charged surface repels approaching ions and suppresses the collected flux, whereas a negatively charged surface accelerates them inward and enhances it. For a Maxwellian ion population with temperature and zero-potential current density given respectively by $T_i, J_{i0}$ the collected ion current density as a function of surface potential is [6]:

$$J_i(\Phi) = \begin{cases} J_{i0} \exp\left(-\frac{e\Phi}{k_B T_i}\right), & \Phi > 0 \\ J_{i0} \left[1 - \frac{e\Phi}{k_B T_i}\right], & \Phi < 0 \end{cases}$$

### 1.5 Triboelectrification

On the lunar surface, triboelectrification is typically induced by human activity or rover operations [29]. To date, there is no simple universal equation that accurately describes all charge-transfer processes involved in triboelectrification; many theories have been proposed, but none is fully established [30].
One of the main parameters determining triboelectrification is the work function difference [30]. When two metals with different work functions, $\Phi_1$ and $\Phi_2$, come into contact, a voltage difference arises at the interface, which is referred to as the contact potential difference (CPD). Due to this CPD, electron transfer occurs between the two metals as follows [31]:

$$V_{CPD} = -\frac{\Phi_1 - \Phi_2}{e}$$

where $V_{CPD}$ is the CPD of metal 1 with respect to metal 2.
Insulator–metal triboelectrification can be described analogously to the metal–metal case, namely in terms of the CPD arising from differences in surface electronic energy levels. Therefore, the work function of an insulator can be defined as an effective work function through triboelectrification experiments using a metal with a known work function. This quantity is often referred to as the effective work function of the insulator. In this case, the CPD is expressed as [30]:

$$V_{CPD} = -\frac{\Phi_I - \Phi_M}{e}$$

where $\Phi_I$ is the effective work function of the insulator and $\Phi_M$ is the work function of the metal.
Because triboelectrification arises from contact between surfaces, the geometry and extent of the contact interface strongly influence charge transfer. When two objects come into contact, surface roughness and morphology prevent the entire apparent area from making true contact. Electron transfer can nevertheless occur by tunneling across very small gaps, with a threshold distance of approximately 2 nm having been suggested [32]. The effective contact area may therefore be regarded as the portion of the opposing surfaces separated by about 2 nm or less. Applied pressure or particle collisions increase this area by reducing the interfacial separation and deforming the surfaces, thereby enhancing charge transfer [33]. For the same reason, triboelectrification can occur even between particles of the same material when their grain sizes differ [34], because differences in grain curvature alter the effective contact area between them. This shows that surface morphology can influence both the magnitude and direction of electron transfer.

## 1.6 Lunar Regolith Properties

Because regolith properties—including mineralogy, grain-size distribution, topography, and electrical characteristics—vary across the Moon, different regions can develop different surface potentials even under identical external charging conditions. This section describes the chemical and physical properties of lunar regolith that are relevant to lunar surface charging.

### 1.6.1 Chemical Properties

Lunar regolith is composed primarily of silicate and oxide minerals formed under anhydrous, chemically reducing conditions. Unlike terrestrial soils, lunar material is essentially free of hydrated minerals such as clays, micas, and amphiboles, and iron exists predominantly in the elemental ($Fe^0$) and ferrous ($Fe^{2+}$) oxidation states rather than the ferric ($Fe^{3+}$) state common on Earth [35]. Furthermore, long-term exposure to the harsh space environment affects the physical properties of regolith grains. Micrometeorite bombardment and solar wind sputtering produce submicroscopic, so-called nanophase, metallic iron (np-$Fe^0$) inclusions and glassy agglutinate coatings on grain surfaces through space weathering [36]. These fundamental chemical differences have direct consequences for the electrostatic behavior of the surface, because they control the work function, secondary electron yield, and dielectric response of the regolith.

Apollo sample analyses indicate that the major oxides—$SiO_2$, $Al_2O_3$, FeO, MgO, CaO, and $TiO_2$—collectively account for more than 98 wt.% of the bulk lunar regolith [35, 37]. The relative proportions of these oxides vary systematically between the two principal terrain types. Feldspathic highland soils are characterized by higher $Al_2O_3$ (24.47 wt.%) and CaO (14.65 wt.%) contents, reflecting the dominance of calcium-rich plagioclase feldspar (anorthite, $CaAl_2Si_2O_8$). Mare basalt soils are characterized by higher FeO (≈16 wt.%) and $TiO_2$ (2–8 wt.%) contents, reflecting the greater abundance of mafic minerals such as pyroxene, olivine, and the iron–titanium oxide ilmenite ($FeTiO_3$) [35, 37]. Table 1 lists representative oxide compositions for highland and mare regolith, compiled from Apollo sample data in the Lunar Sourcebook [35] and the NASA Lunar Regolith Simulant User's Guide [38].

*Table 1. Representative oxide compositions (wt.%) of highland and mare regolith from Apollo samples [38].*

| Apollo Bulk Chemistry (wt.%) | | | | | | | | |
|---|---|---|---|---|---|---|---|---|
| Region | $SiO_2$ | $TiO_2$ | $Al_2O_3$ | FeO | MgO | CaO | $Na_2O$ | $K_2O$ |
| Highlands | 45.62 | 0.84 | 24.47 | 6.6 | 6.98 | 14.65 | 0.48 | 0.2 |
| High-Ti Mare | 41.27 | 7.48 | 12.5 | 16.47 | 9.75 | 11.26 | 0.39 | 0.11 |
| Low-Ti Mare | 45.62 | 2.09 | 13.2 | 16.45 | 10.19 | 10.65 | 0.42 | 0.19 |

### 1.6.2 Physical Properties

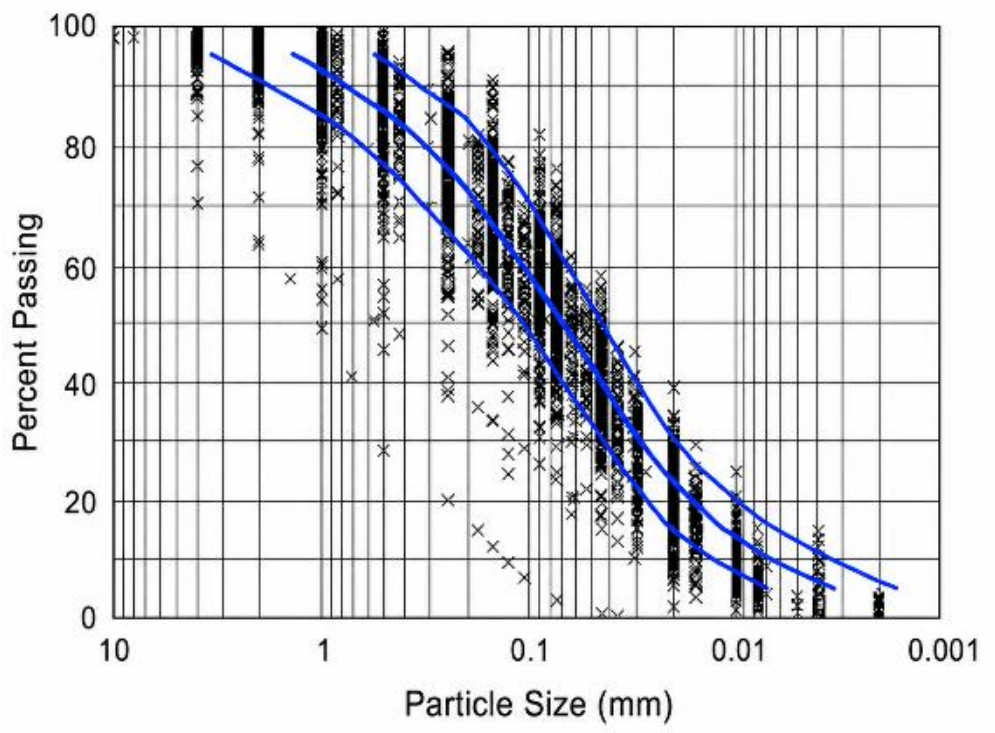

*Figure 12. Particle size distribution for all Apollo returned samples, showing that lunar soils consist of fine-grained particles spanning a broad range of sizes [38].*

Lunar regolith covers a wide range of particle sizes, from boulders down to submicron dust. The bulk fine-grained soil has a median grain size of roughly 40–100 µm, with a significant portion below 20 µm [36, 38]. Figure 12 shows the particle size distributions of Apollo soils and illustrates that lunar regolith contains abundant fine-grained particles. This abundance of fine grains contributes to the high surface-area-to-volume ratio, making them particularly sensitive to surface charging. Grains are typically angular and irregular due to impact fragmentation. The regolith has a loosely packed, porous structure with many internal voids that can trap particles and charges.

The current-balance condition of Section 1.1 applies equally to an individual dust particle above or within the sheath [9]. For a spherical grain of radius $a$, the equilibrium grain potential $\Phi_g$ is obtained by requiring that the integrated currents over the grain surface sum to zero:

$$\sum_k I_k\left(\Phi_g\right) = 0, \qquad I_k = 4\pi a^2 J_k$$

Here $I_k$ is the total current from mechanism $k$ to the grain. The individual current expressions are modified to account for the finite grain size relative to the Debye length and for anisotropic illumination and plasma flow [10, 39]. Once the equilibrium charge $Q = C_g\Phi_g$ is known, with $C_g = 4\pi\varepsilon_0 a$ the grain capacitance, the electrostatic force $QE$ can be compared with lunar gravity $mg_{\mathrm{Moon}}$. When $QE \gtrsim mg_{\mathrm{Moon}}$, electrostatic lofting or levitation becomes possible.

The DC conductivity of lunar soil is extremely low and depends strongly on temperature [40]. Laboratory measurements on an Apollo 15 soil sample in vacuum give $\sigma(T) = 6 \times 10^{-18}\ e^{0.0237T}$ where σ is in S $m^{-1}$ and T is in kelvin [41]. The dielectric constant ε controls how electric fields penetrate the material and how charge distributes near the surface. Lunar regolith values are ε ≈ 2–4, increasing with density and therefore reflecting the degree of compaction [42]; the dielectric response also depends on vertical stress, so in situ values are not fixed by composition alone [43]. Together, ε and σ set the charge relaxation time τ = ε/σ. Applying the relation above gives a relaxation time of order minutes at equatorial daytime temperatures, whereas within a permanently shadowed region it lengthens to roughly 20 days, comparable to one lunation [19].

The threshold for photoemission is set by the energy separation between the valence band maximum and the vacuum level, and it therefore fixes the cutoff wavelength introduced in Section 1.2.1. For Apollo lunar fines this threshold has been measured at approximately 5 eV [15, 16]. It varies between minerals: scanning-probe measurements give 4.29 eV for ilmenite, 5.14 eV for pyroxene, 5.58 eV for plagioclase, and 7.90 eV for olivine [44].

A related but distinct quantity governs contact electrification. Because an insulator has no well-defined Fermi level, its contact-charging behavior is instead described by an effective work function, determined by measuring charge transfer against metals of known work function (Section 1.5). For the lunar simulant JSC-1, this effective work function is approximately 5.8 eV [45]. Although similar in value, it should not be equated with the photoemission threshold, as the two represent distinct physical concepts governed by different mechanisms.

## 1.7 Variations Across Time

### 1.7.1 Diurnal Variation

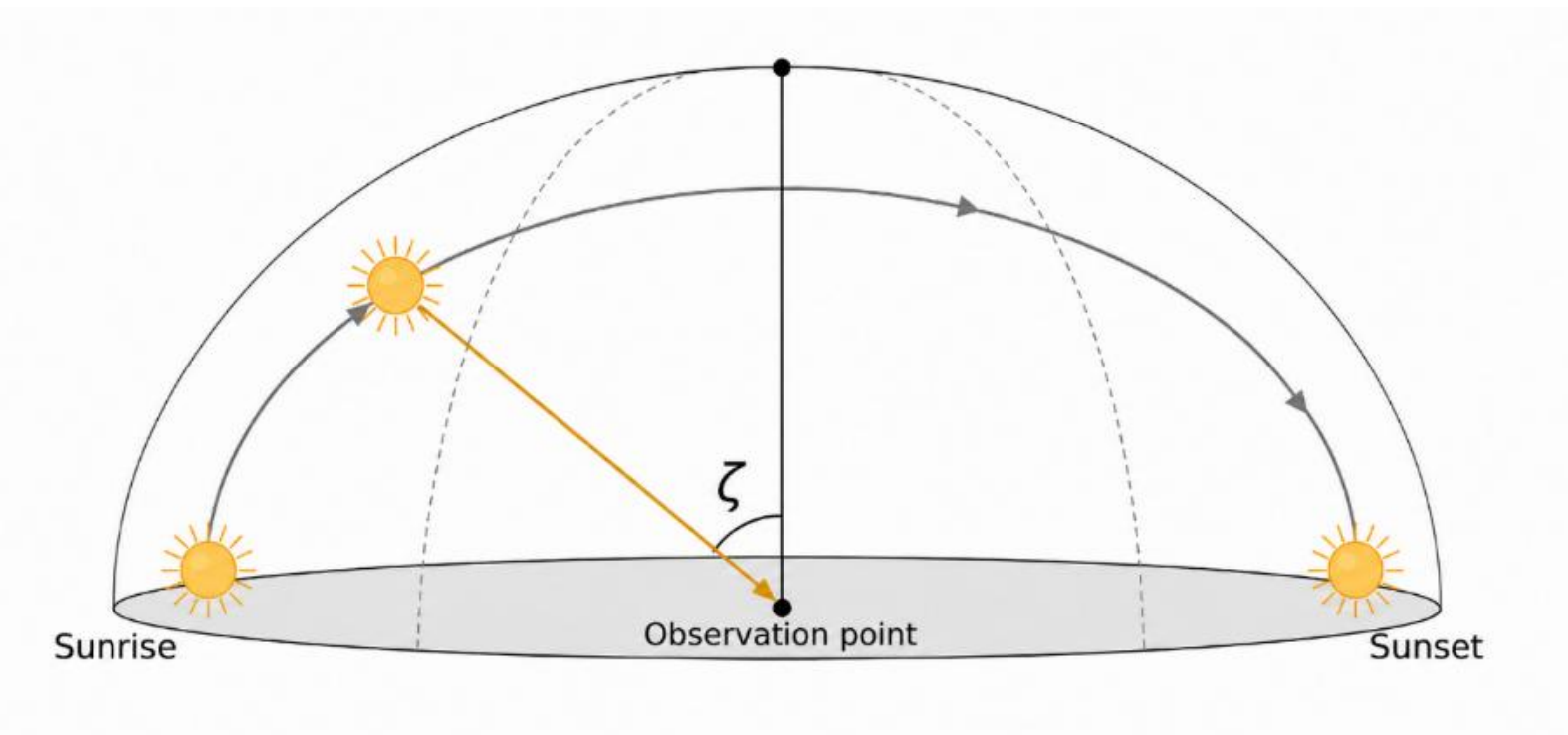

*Figure 13. Lunar diurnal cycle and solar zenith angle.*

Because the Moon's rotation is synchronous with its orbit about Earth, each surface location experiences a solar day of approximately 29.5 Earth days, progressing from lunar noon through sunset, the long lunar night, and sunrise. Accordingly, the solar zenith angle and all sun-dependent inputs such as illumination, solar wind, SEP and temperature vary along the lunar day.

Illumination geometry introduces a strong spatial modulation of the photon environment. As illustrated in Figure 13, the solar zenith angle ζ is defined as the angle between the incident solar direction and the local zenith at the observation point. The incident energy flux is proportional to the cosine of the solar zenith angle, cos ζ, so both the total irradiance and the photon flux decrease as the solar zenith angle increases.

A shadow boundary separates two regions with different current balances: photoemission ceases within the shadow while the ambient plasma-electron current continues, so the two sides settle at different potentials [7]. The resulting gradient is steepest near the terminator, where a small change in solar zenith angle sweeps a large change in shadow geometry. A moving shadow boundary can also drive the surface away from equilibrium altogether. Because the surface charging time introduced in Section 1.6.2 is finite, a terminator that sweeps across a region faster than the local potential can re-equilibrate leaves the surface in a transient state: the newly darkened area stops emitting photoelectrons but continues to collect them from adjacent illuminated surfaces, and those illuminated surfaces consequently lose electrons faster than they recover them, transiently reaching more than twice their equilibrium positive potential [14]. This transient enhancement is referred to as the supercharging effect.

Diviner Lunar Radiometer measurements show that equatorial daytime maximum temperatures reach approximately 387–397 K, whereas predawn minimum temperatures fall to approximately 95 K [46]. These large temperature variations can modify the charging-related material properties discussed in Section 1.6, such as electrical conductivity, and thereby affect the charging state of the lunar regolith.

### 1.7.2 Monthly

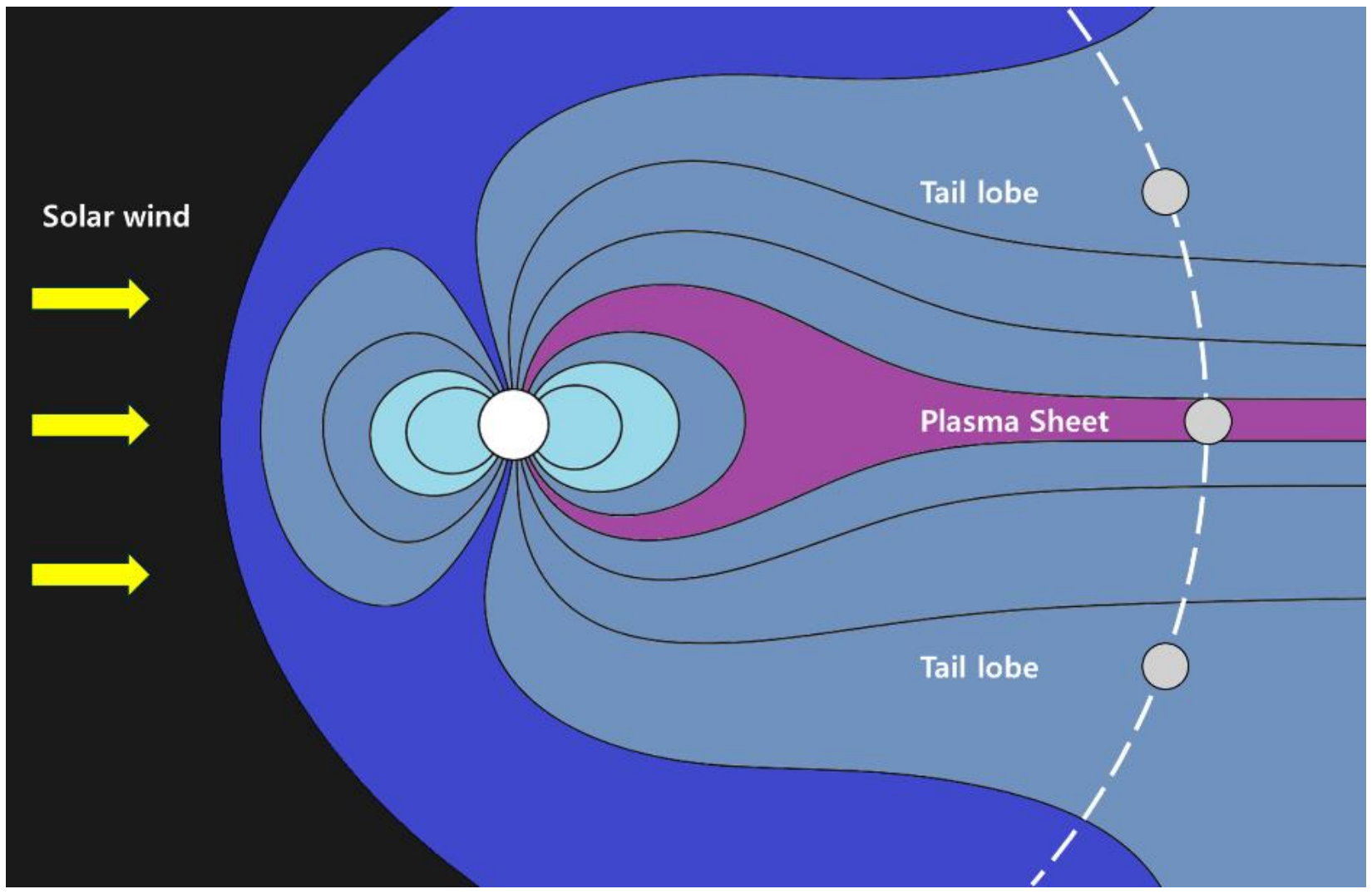

*Figure 14. Moon's passage through Earth's magnetotail.*

Because the Moon orbits Earth, it passes through Earth's magnetotail approximately once each lunar month, as illustrated in Figure 14. As discussed in Section 1.4, the lunar surface is exposed primarily to solar-wind plasma outside the magnetotail, whereas during magnetotail passage it encounters plasma with substantially different properties. Plasma conditions also vary within the magnetotail depending on whether the Moon is located in the tail lobe, plasma sheet boundary layer, or central plasma sheet.

The local magnetic-field geometry also affects charged-particle transport toward the lunar surface. Under the Lorentz force, charged particles gyrate around magnetic-field lines while moving along them in helical trajectories. The magnetic-field orientation relative to the surface can therefore affect which charged particles reach the surface and their incidence directions.

### 1.7.3 Solar Cycle

The approximately 11-year solar cycle produces long-term variations in the solar radiation, solar-wind plasma, and SEP environments reaching the Moon. During solar maximum, EUV and soft X-ray fluxes increase by about 30% or more, while FUV emission, including Lyman-α, increases by approximately 3–25% [47]. The solar magnetic field becomes more complex, the solar wind becomes more variable, and its average density and electron temperature generally increase [3]. ICMEs also occur more frequently and can temporarily raise the plasma density and dynamic pressure to several times their nominal values [48]. SEP events are more frequent, and their cumulative fluence is greater, increasing the occurrence of intense charging events [49]; nightside potentials as negative as approximately −4.5 kV have been reported during large SEP events [17, 50].

During solar minimum, the solar UV output, average solar-wind density, and electron temperature are generally lower. The large-scale solar magnetic field is more ordered, producing more clearly separated fast polar and slow equatorial solar-wind streams, with fast streams from coronal holes becoming more prominent [3]. SEP events occur less frequently and their total fluence rate decreases. However, the mean fluence per event tends to increase as solar activity declines, indicating that individual extreme events may still occur near solar minimum despite their lower frequency [49].

Conversely, the GCR flux varies inversely with solar activity. Near solar maximum, the stronger heliospheric magnetic field limits the penetration of GCRs into the inner solar system. Passing ICMEs and their associated shocks can also produce short-term reductions in GCR flux, known as Forbush decreases [51]. Consequently, the GCR flux at the Moon decreases from approximately 4 particles $cm^{-2}$ $s^{-1}$ near solar minimum to about 2 particles $cm^{-2}$ $s^{-1}$ near solar maximum [19].

# 2 Energetic Particle Penetration and Deep Dielectric Charging

Among the primary energetic-particle sources described in Section 1.3, SEPs and GCRs are the principal populations capable of penetrating beyond the immediate surface layer and producing charge-generating interactions within the lunar subsurface [19, 52]. Solar-wind electrons and protons possess relatively low characteristic energies of approximately 1 keV, resulting in nanometer-scale penetration depths and therefore a limited direct contribution to deep subsurface charging [53]. GCRs are capable of penetrating tens of centimeters into the regolith [54, 55], whereas SEPs may penetrate up to a few millimeters [52]. These values represent characteristic transport scales rather than unique stopping depths, because individual particles undergo stochastic sequences of scattering and energy-loss interactions. The accessible interaction depth depends on particle species and incident energy, together with the stopping properties and structure of the regolith. Penetration therefore determines where energy transfer and charge-carrier generation may occur, but it does not by itself determine the resulting net-charge distribution. Because the anhydrous lunar regolith acts as an excellent electrical insulator, residual uncompensated or spatially separated charges may dissipate slowly and form depth-dependent charge distributions capable of generating internal electric fields [52].

## 2.1 Stochastic Particle Transport

At the macroscopic level, the transport of energetic particles through matter can be described within the framework of the linear Boltzmann transport equation, which accounts for particle motion, scattering, absorption, energy loss, and secondary-particle production. Monte Carlo methods represent this transport by sampling individual particle histories from the probability distributions associated with the relevant interaction processes [56]. When an energetic charged particle enters the regolith, its trajectory is governed by a stochastic sequence of interactions rather than a single deterministic collision path [56, 57, 58]. The interaction probabilities depend principally on the particle species and kinetic energy, along with the atomic composition and atomic number density of the target material [59, 60]. In a mineralogically heterogeneous regolith, the local interaction probability changes as the particle passes through different mineral grains and glassy components. At the bulk scale, these local variations may be approximated by using composition-weighted, mixture-averaged macroscopic cross sections [59]. Such a homogenized approximation provides an effective description of composition-averaged interaction probabilities, but it does not resolve individual grain boundaries, pore geometry, neighboring grain interactions, or local electrostatic potentials. These unresolved microscale properties become particularly important when determining how generated and incident charges are transported, redistributed, and ultimately retained within the regolith. In a Monte Carlo description, the distance s to the next discrete interaction is sampled from the total mean free path, while the interaction type is selected according to the relative macroscopic cross section of each available process:

$$s = -\lambda_{tot}(E)\ln\varepsilon_1$$

$$P_i(E) = \frac{\sum_i(E)}{\sum_{tot}(E)}$$

where $\varepsilon_1$ is a uniformly distributed random number, $\lambda_{tot}$ is the total mean free path, and $P_i$ is the conditional probability that the discrete interaction occurs through process i. Different values of $\varepsilon_1$ produce different free-flight distances even for particles with the same initial energy [59, 60]. At the sampled interaction point, the process is selected according to $P_i(E)$, while the scattering angle, transferred energy, and any secondary-particle properties are subsequently sampled from the corresponding final state distributions [59, 60]. Consequently, two particles with identical initial energies and incidence angles may undergo different sequences of elastic and inelastic events, producing different trajectories, stopping depths, energy-deposition profiles, and primary-particle outcomes [58, 60]. Having established how the location and type of an interaction are selected, the next step is to determine how each process redistributes the momentum and energy and how these transfers influence the outcomes of the primary particle.

## 2.2 Energy Transfer Pathways and Primary-Particle Outcomes

Once an interaction has been selected, its immediate consequence depends on how momentum and energy are transferred. At the microscopic level, charged-particle interactions may be classified as elastic and inelastic [60, 61].

For incident electrons, elastic scattering occurs through interactions with the atomic potential and primarily changes the direction of the electron [60]. Since the target atom is much more massive than the incident electron, the energy transferred to atomic recoil during an individual elastic event is generally small compared with the resulting angular deflection [60]. Therefore, elastic scattering has a primary role in determining the geometry of the electron trajectory [60].

Inelastic electron interactions instead transfer kinetic energy to the target [60]. If the transferred energy is sufficient for excitation but below the ionization threshold, electronic excitation may occur, where a target electron gets promoted to a higher energy state without being removed from the material. If the energy exceeds the binding energy, ionization may occur, ejecting a target electron and thus leaving a positively ionized site or hole [60]. At sufficiently high incident energies, radiative losses through bremsstrahlung may also contribute, although excitation and ionization are more directly involved in the local generation and redistribution of charge within the regolith [59, 60, 61].

Proton interactions follow an analogous transport sequence but involve different energy-transfer channels. Coulomb interactions with target electrons produce excitation and ionization. The average energy lost by the proton through these interactions per unit path length is described by the electronic stopping power [61].

Elastic collisions with target nuclei transfer energy to recoiling atoms, potentially deflecting the proton and displacing target atoms. The corresponding average energy loss per unit path length is described by the nuclear stopping power [61]. Thus, the total proton stopping power is the sum of the electronic and nuclear contributions [61]. These quantities describe how the proton transfers and loses kinetic energy but do not determine whether the proton stops or exits the selected control volume, or whether its positive charge is retained within that volume [61, 62].

As these interactions accumulate, the primary particle may reach one of three broad outcomes with respect to a specified control volume. It may lose energy and stop within the volume, with a proton becoming implanted or an electron coming to rest. It may be redirected through the incident boundary by backscattering, or it may exit through another boundary by transmission [60, 62]. These outcomes determine the final states of the primary particles, but they do not necessarily describe the complete energy or charge balance of the interaction. A backscattered or transmitted primary particle may deposit substantial energy before leaving, whereas a stopped or implanted particle may remain even though secondary electrons generated along its trajectory escape [62].

The particle-resolved simulations of Gandhi et al. [62] provide an illustrative example of this. In one simulated trajectory, a 1 keV solar-wind proton deposited approximately 950 eV in $SiO_2$ before backscattering and leaving the simulation volume. Most of its initial kinetic energy was transferred to the material, although the positive charge of the primary proton was not retained within the selected volume. By contrast, simulated protons that became implanted retained their incident positive charge in addition to transferring energy to the target.

Primary-particle energy deposition and primary-charge retention must therefore be treated as physically distinct outcomes [62]. This model involves keV solar-wind particles interacting with idealized homogeneous amorphous $SiO_2$ grains and should not be interpreted as a material analog of lunar regolith under SEP or GCR irradiation. The heterogeneous nature of lunar regolith may alter the stopping depths, energy-loss distributions, and the probabilities of implantation, backscattering, and transmission. This simulation is therefore useful for illustrating the general charge-accounting principle that substantial energy deposition may occur even when the incident primary charge is not retained within the selected control volume.

The charging consequence of a primary particle outcome is also boundary dependent. A primary particle that exits the grain of origin may enter and stop within a neighboring grain, thereby altering the grain-scale charge distribution without leaving a bulk control volume containing both grains. A complete charge balance additionally requires consideration of the secondary electrons generated during inelastic interactions and whether they remain within the material, escape from the grain of origin, become reabsorbed elsewhere in the regolith, or leave the selected bulk control volume.

## 2.3 Secondary Electron Generation and Escape

Secondary electrons are target electrons generated when incident particle interactions excite or ionize the material [25, 63]. Secondary electron emission (SEE) may be described as a sequence of three physical stages: the generation of excited electrons through primary-particle interactions, their transport toward the solid-vacuum interface, and their escape across the material surface barrier [63]. Generation alone does not guarantee emission. Electrons produced too far below the surface or those that lose sufficient energy through subsequent interactions may thermalize or recombine before reaching the surface. Some carriers may also become trapped in material-dependent localized electronic states before reaching the surface. The possible contribution of such trapping is discussed in Section 2.6. Only a fraction of the generated population reaches the surface with sufficient energy and an outward trajectory to cross the material surface barrier [25, 63].

The material surface barrier should be distinguished from the electrostatic potential established by previously accumulated charge. Crossing the solid-vacuum interface means that an electron has been emitted from the material, but it does not mean that the electron will escape from the charged grain. If the grain or emitting surface patch is at a positive potential relative to its surroundings, the emitted electron is electrostatically retarded and may be drawn back to the surface [25]. Only electrons whose outward kinetic energy exceeds the corresponding electrostatic potential-energy difference can escape rather than be recollected. By contrast, a surface at a negative potential relative to its surroundings generally accelerates an emitted electron away from itself, reducing the probability of recollection by the grain of origin. This electrostatic acceleration does not increase secondary-electron generation, or guarantee escape from packed regolith, since local pore geometry and the potentials of neighboring grains may redirect or reabsorb the emitted electron [25].

Escape from the grain of origin must also be distinguished from escape from the bulk regolith. The charging consequence depends on the spatial scale at which escape occurs. Three spatial boundaries are considered here: crossing the surface of the grain of origin, transport through a local pore or grain cluster, and complete escape across the outer boundary of a selected bulk regolith control volume. An electron that crosses the surface of one grain may enter a narrow pore or microcavity whose surrounding solid angle is largely occupied by neighboring grain surfaces. The emitted electron may then intersect and be reabsorbed by another grain or pore wall rather than leave the regolith. In this case, the emitting grain becomes more positive and the receiving grain more negative, producing inter-grain charge separation even though no electron has been lost from a control volume containing both grains [62, 64]. An electron produces a net loss of negative charge from the selected control volume only if it crosses the outer boundary of that volume. Grain-scale emission, pore-scale redistribution, and bulk-regolith escape must therefore be treated as physically distinct outcomes. The combined contributions of the primary particle, generated secondary electrons, and ionized target sites are considered in the following section.

## 2.4 Single Interaction Charge Retention and Redistribution

The retained-primary-electron outcome described in the preceding section is commonly represented in grain charging models as electron sticking. If an incident electron loses its kinetic energy and remains associated with the grain, it contributes negative charge, whereas secondary-electron emission removes negative charge from the grain and shifts the grain charge in the positive direction [65]. Sticking and emission are therefore competing grain-scale charging pathways.

However, neither process alone determines the charge retained within a bulk regolith control volume. As described in Section 2.3, an emitted electron may be recollected by the original grain, reabsorbed elsewhere in the regolith, or lost from the specified control volume. An electron generated by ionization may also recombine with its nearby positive site, preventing persistent charge separation [25, 62, 63, 64]. After these processes, the residual positive and negative charges may remain on different grains and at different locations within the regolith.

Since these residual charges may be spatially separated, the outcome of one particle interaction cannot always be described by a single net retained-charge value. Here, the retained-charge distribution refers to the spatial distribution of positive and negative charge remaining within a specified control volume after primary particle transport, secondary electron transport, reabsorption, and recombination. It may include the charge of a stopped or implanted primary particle, electron-hole separation, and charge separation produced by electron transfer between neighboring grains. This distribution should be distinguished from the net retained charge, which refers to the total residual charge within the control volume. For example, electron transfer between neighboring grains produces local charge separation without changing the net charge of a volume containing both grains [62, 64].

For a single incident particle, the residual charges may remain on different grains and at different depths below the bulk-regolith surface. Since particles with identical initial conditions may undergo different stochastic interaction pathways [58, 59, 60], a macroscopic charging model would require the average signed charge remaining at each unit depth per incident particle, distinguished by particle species and incident energy. If this depth-dependent retained-charge response were known, it could be combined with the corresponding incident particle flux to estimate the volumetric rate at which charges are supplied as a function of depth. This connection relies on three assumptions: grain-scale outcomes can be represented by a depth-dependent average; the retained-charge responses of successive particles can be combined linearly; and the initial retained-charge distribution is established by particle transport, reabsorption, and recombination before it is modified by slower conduction and dielectric relaxation. These assumptions may become less accurate as accumulated charge modifies local electrostatic potentials, thereby affecting the charge escape, recollection, and redistribution.

As discussed in Section 2.2, energy transfer does not necessarily result in charge retention within the specified control volume. Primary particles may deposit energy before leaving the volume, generated electrons may recombine, and secondary electron transport may relocate charge away from the initial energy-transfer location [59, 60, 62]. The retained-charge distribution therefore cannot be inferred directly from the energy deposition profile. Although the individual interaction mechanisms are known, their combined effects on recombination, reabsorption, trapping, and bulk escape have not been constrained under lunar regolith conditions [62, 64, 66]. The average depth-dependent retained-charge response per particle is therefore a required but unresolved microscopic input to macroscopic charging models.

## 2.5 Effective Penetration and the CSDA Range

The particle pathways described above introduce a distribution of trajectories and interaction locations rather than a single deterministic penetration depth. For bulk-scale estimates, a characteristic transport scale may be estimated using the continuous slowing down approximation (CSDA), in which stochastic energy losses are represented by the average rate of kinetic energy loss within the target material [67].

The CSDA mass range of an electron with initial energy $E_0$ is given by the integral of the reciprocal mass stopping power. When converted using an assumed bulk density, an equivalent linear range may be written as:

$$z_{CSDA} = \frac{1}{\rho_{bulk}} \int_{0}^{E_0} (S_{mass}(E))^{-1} dE$$

The CSDA represents the stochastic energy losses experienced along an electron trajectory as a continuous average loss until its kinetic energy is exhausted. The resulting linear CSDA range is represented by $z_{CSDA}$, where $E_0$ represents the incident energy, $\rho$ is the bulk density of the material, and $S_{mass}(E)$ represents the total mass stopping power of the medium. For electrons, the total stopping power includes collisional energy loss through excitation and ionization and radiative energy loss via Bremsstrahlung emissions. The equation shows that the resulting linear CSDA range increases with the incident energy and is inversely proportional to bulk density of the regolith.

However, the CSDA range should not be interpreted as the unique physical depth at which an electron stops. It approximates the average path length traveled while the particle loses its energy, whereas the projected penetration depth is measured along the initial direction of incidence and is reduced by multiple scattering. The CSDA neglects energy-loss straggling and does not explicitly resolve backscattering, transmission across finite grain boundaries, or the spatial distribution of deposited energy [67, 68]. It therefore provides a useful characteristic scale for primary-electron transport but does not directly determine the depth profile of energy deposition or the retained-charge distribution.

## 2.6 Charge trapping, Relaxation, and Internal Electric Field generation

The particle transport processes described in the preceding sections determine where energy is transferred and charge carriers are generated. However, not all generated carriers contribute to persistent subsurface charging.

Following primary-particle stopping or implantation, secondary electron emission, reabsorption, and recombination, a residual distribution of uncompensated or spatially separated charge may remain in the lunar regolith. This residual charge distribution provides the microscopic basis for charge accumulation.

As elucidated in Section 2.4, the microscopic input required to connect these individual interaction outcomes to a macroscopic charging model is the average signed charge retained per unit depth per incident particle, distinguished by particle species and incident energy. If this retained-charge outcome were known, it could be combined with the incident particle flux to estimate the rate at which positive and negative charge are supplied per unit volume at each depth. Repeated irradiation could then form a macroscopic subsurface charge distribution that accumulates over time. This microscopic to macroscopic transition is summarized in Figure 15

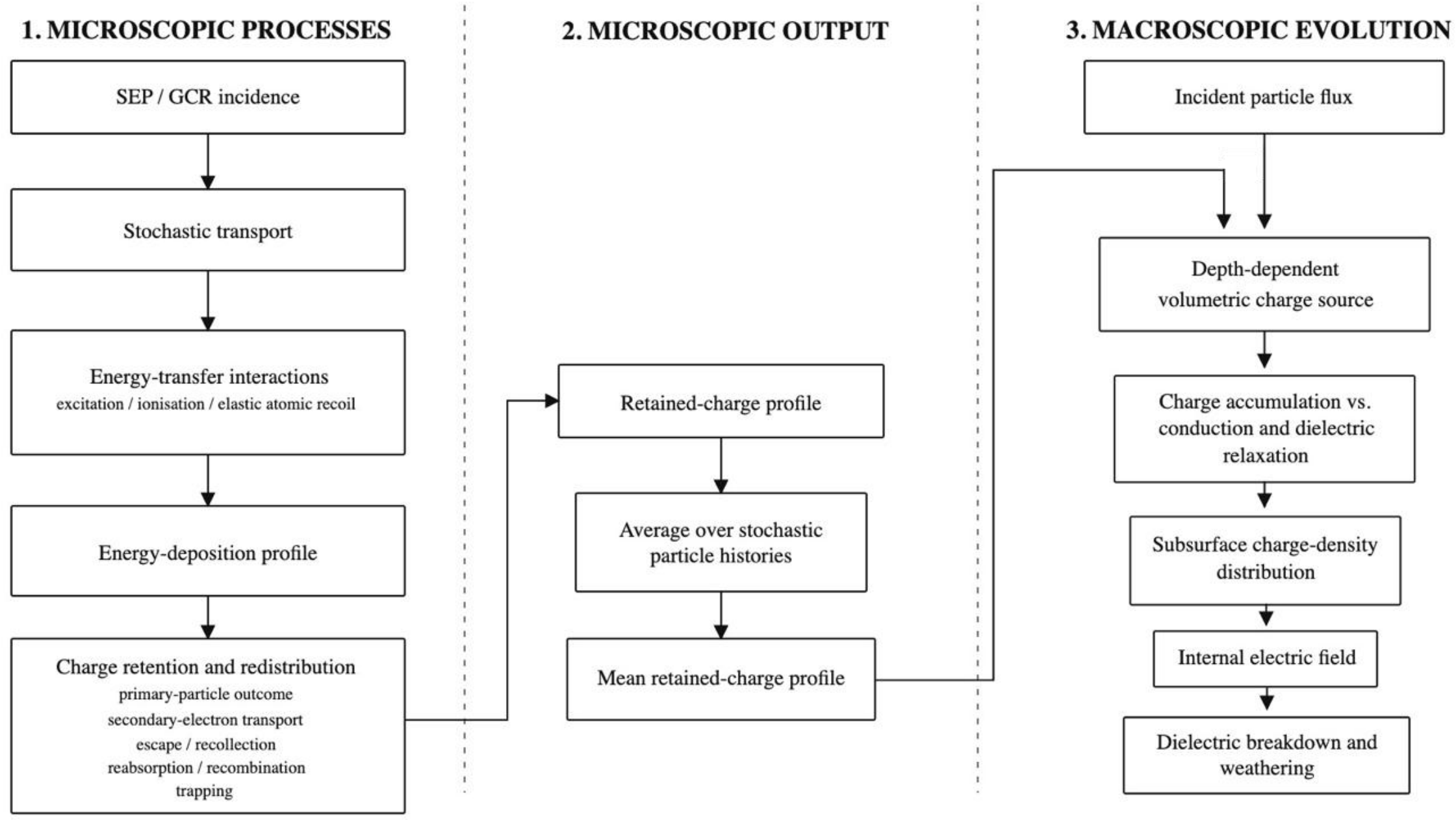

*Figure 15. Conceptual framework linking microscopic energetic-particle interactions to macroscopic subsurface charging.*

Microscopic trapping may contribute to the persistence of this residual charge. Studies of dielectric materials show that crystalline defects, radiation-induced vacancies, and amorphous materials may create localized electronic states that restrict charge-carrier transport [62, 69]. Such states may capture electrons or holes and reduce their effective mobility, although the extent of carrier localization is material-dependent and cannot be inferred from structural disorder alone [69, 70]. Given the abundance of impact-generated glass and radiation-processed material in lunar regolith, defect-assisted localization may contribute to the persistence of subsurface charge [19, 71, 72]. However, their actual contribution to charge trapping under lunar regolith conditions remains unknown. Microscopic trapping should be considered a plausible but unquantified process and should be distinguished from the bulk relaxation of the resulting charge distribution.

The persistence of the resulting subsurface charge distribution can be characterized at the bulk scale by the dielectric relaxation time:

$$\tau = \frac{\varepsilon}{\sigma}$$

where $\tau$ stands for the charge relaxation timescale, $\varepsilon$ represents the absolute dielectric permittivity, and $\sigma$ represents the bulk electrical conductivity of the material. Charge can accumulate when energetic-particle irradiation supplies retained charge more rapidly than it can be redistributed or dissipated. In cryogenic PSRs, reduced bulk conductivity is expected to increase the dielectric relaxation time substantially [40]. Under PSR conditions, relaxation times may extend to several weeks, allowing residual charge to persist after the incident flux has declined [19]. Energetic-particle penetration determines where charge-generating interactions may occur, whereas microscopic trapping and dielectric relaxation determine what charge remains and how long it persists. Modeling the resulting internal electric field therefore requires the spatial and temporal evolution of the separated positive and negative charge distributions.

To model the consequences of charge accumulation within PSRs, Jordan et al. [19] proposed a one-dimensional, two-layer, time-dependent model. This model simplifies the volumetric penetration of particles by partitioning the incoming fluxes into two layers of non-overlapping, infinitely planar current densities composed of a shallower positive current along with a deeper negative electron current. The separation of these layers generates an internal electric field. This simplification allows the subsurface to be treated as a parallel plate capacitor, where the internal electric field is governed by the net charge differential across the gap between the layers. The subsurface conductivity then acts to gradually dissipate the charge over time [19, 71].

By applying Gauss's law to this model, Jordan calculated the electric fields created from the two layers by incorporating CRaTER and ACE/EPAM flux data to estimate the subsurface electric fields within lunar PSRs from SEP and GCR events. Jordan et al. [19] report that GCRs produce persistent fields up to ~700 V/m whereas large SEP events produce transient fields of >$10^6$ V/m. The simulation results showed that large SEP events can deposit charge faster than the regolith can dissipate it, suggesting that SEP events may drive the electric field above the threshold for dielectric breakdown and contribute to the weathering of lunar regolith, which will be articulated further in the next section [19, 71].

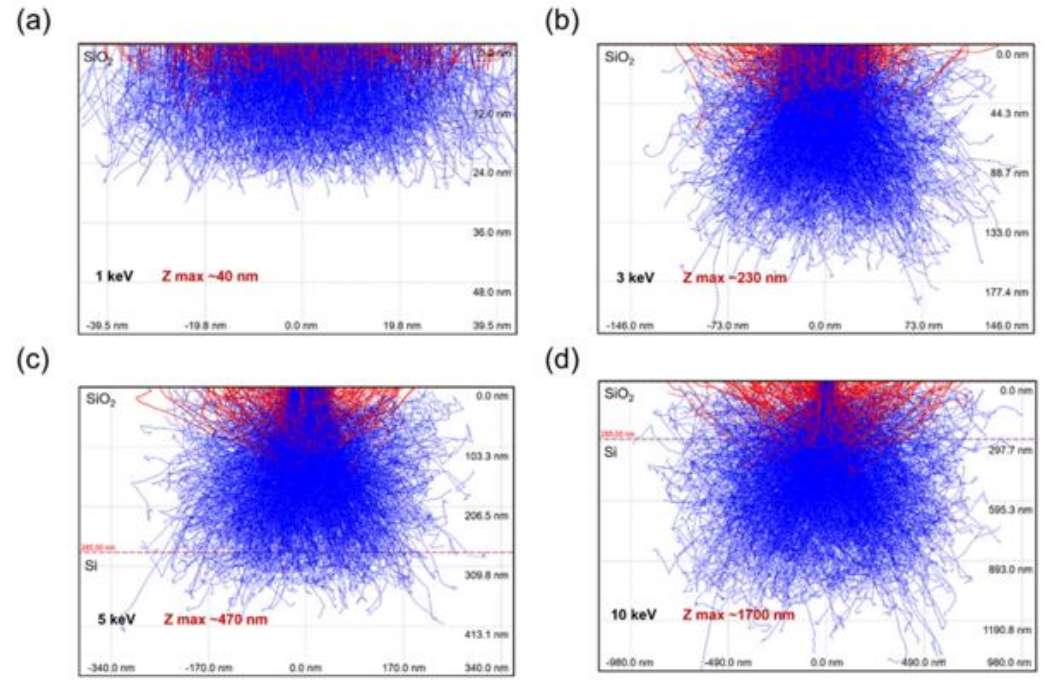

*Figure 16. Monte Carlo simulations of electron trajectories in $SiO_2$ at incident energies of 1 keV, 3 keV, 5 keV, and 10 keV [73].*

While this 1D parallel plate model establishes the principles of internal electric field generation, it uses a geometric idealization by assuming that particle-induced charging can be represented by planar layers. In reality, incident electrons and protons do not stop along uniform lines. Instead, stochastic scattering and energy loss interactions occur throughout three-dimensional interaction volumes [66].

As illustrated by Monte Carlo electron trajectory simulations (Figure 16), increasing the incident electron energy produces broader, three-dimensional interaction volumes in $SiO_2$ as elastic and inelastic scattering redistribute the electron trajectories [73]. The figure therefore illustrates that particle interactions occur over finite depth ranges rather than along planar interfaces. As mentioned in Section 2.2, the $SiO_2$ simulation is used to illustrate general transport geometry and is not treated as a direct analog of lunar regolith.

Sana et al. [66] represented the incident proton and electron currents and the resulting charge accumulation as continuous functions of depth rather than discrete planar layers. Furthermore, during SEP events, both protons and electrons can penetrate the lunar surface, causing these charging zones to overlap. Hence, instead of non-overlapping, planar sheets, overlapping regions of positive and negative volumetric charge densities can develop within the lunar subsurface [52, 66]. Compared with the Jordan model, this approach replaces the two planar layers with a continuous, depth-dependent representation of particle-induced charging.

To represent volumetric penetration, Sana modeled the depth-dependent particle currents and coupled the charge continuity equation with Poisson's equation to estimate subsurface electric fields within lunar PSRs under SEP and GCR irradiation [66]. The model predicts GCR-driven persistent fields from ~115 V/m during solar maximum to ~320 V/m during solar minimum, whereas large SEP events generate transient but intense subsurface fields around $10^6 – 10^7$ V/m [66]. Since the low temperatures of the PSRs restrict internal charge drainage, these localized field gradients may persist beneath the surface for days.

Sana et al. [66] examined electric field profiles as a function of depth and identified specific locations where the electric field exceeded dielectric breakdown thresholds. The model suggests that the locations of the peak electric fields are not determined solely by particle fluence, but rather by the volumetric charge density within the lunar subsurface [66]. This study highlights the importance of considering volumetric charge distributions when assessing electric field generation and dielectric breakdown processes in the lunar environment. The results also suggest that subsurface charging should not be regarded as an isolated process. Since volumetric charge accumulation generates electric fields that may extend toward the surface, a broader perspective that considers both surface and subsurface charging processes may be necessary to fully characterize the lunar electrostatic environment. Future studies should investigate whether subsurface charge distributions are able to influence near-surface electric fields and modify surface charging behavior.

Jordan et al. [19] represent particle-induced charging using two-layer, planar current sheets, whereas Sana et al. [66] represent the particle currents and resulting charge density continuously with depth. The Sana model therefore provides a more spatially resolved macroscopic description compared to the Jordan model. However, neither model derives its macroscopic charge source from the average signed retained-charge response per incident particle. Neither model calculates the charge that remains after primary particle escape, secondary electron transport, reabsorption, recombination, trapping, and escape from the selected bulk-regolith control volume. The conversion from particle transport and energy transfer to a retained-charge distribution therefore remains unresolved under realistic lunar conditions.

The required microscopic inputs remain insufficiently constrained, including the fractions of generated electrons that recombine, are recollected, are reabsorbed, become trapped, or escape from the bulk regolith. The charge contributions of stopped, implanted, backscattered, and transmitted primary particles must also be included. The dependence of these processes on mineralogy, grain and pore geometry, impact-glass content, and pre-existing electrostatic potential also remains uncertain. Therefore, cryogenic high-vacuum irradiation experiments using

PSR-relevant regolith materials, combined with particle-resolved modeling and measurements of retained charge, electric-field profiles, and charge decay, are required to validate and constrain PSR charging models.
Once the average depth-dependent retained-charge outcome per incident particle has been constrained, it could be combined with SEP and GCR particle spectra to construct a clearer volumetric source term. This source term could then be coupled with conduction, dielectric relaxation, and Poisson's equation to model the evolution of subsurface charge density and internal electric fields. This approach would preserve the macroscopic framework of existing PSR charging models while providing a microscopic physical basis for the charge source used within them.

## 2.7 Dielectric breakdown and Regolith Weathering

Dielectric breakdown is a phenomenon that occurs when a dielectric material experiences a local electric field exceeding its dielectric strength. Consequently, the material loses its insulating properties and transitions into a conductive state which permits rapid current flow by vaporizing small channels within the regolith grain [52, 74]. Fluences of approximately $\sim 10^{10} – 10^{11}$ $cm^{-2}$ and electric fields of approximately $\sim 10^{7}$ V/m for breakdown have been suggested in relevant studies [74, 75, 76, 77]. However, the breakdown threshold of lunar regolith is expected to vary by its composition, microstructure, temperature, and local field conditions.
Within lunar PSRs, dielectric breakdown is of concern due to the low electrical conductivity and long dielectric relaxation time of the cryogenic regolith [19]. While GCRs produce steady subsurface electric fields, episodic SEP events may drive internal electric fields past the regolith's breakdown threshold [19, 52, 66]. One modeling study estimated that gardened regolith within PSRs may have experienced about $10^{6}$ SEP events that are capable of causing dielectric breakdown [66].
Several microstructural characteristics of the lunar regolith enhance local electric fields and reduce the effective dielectric strength of the subsurface. The irregular and angular geometries of the regolith grains can amplify local electric fields. When electric charge accumulates near the sharp grain asperities, modeling studies suggest that local electric fields may be enhanced by 1-2 orders of magnitude [78]. In addition, mineralogical inclusions within the regolith grains introduce dielectric discontinuities because adjacent phases often possess contrasting dielectric properties [79]. The boundaries between these dielectrics can concentrate the local electric field and reduce the resistance of the material to breakdown [80, 81, 82, 83]. Pore spaces and gas-filled inclusions may assist the physical disruption of grains during breakdown because gases normally have a lower dielectric strength than solids [52, 84]. Gas ionization can generate localized microexplosions that induce mechanical stresses on the grains, and repeating these stresses may drive grain fragmentation during breakdown [81]. These mechanisms are supported by general dielectric studies, but their importance under lunar regolith conditions remains unquantified.
Collectively, these features may lower the effective breakdown threshold of lunar regolith and suggest that any process capable of altering grain morphology, porosity, or internal grain structures may influence the likelihood of dielectric breakdown.
Although dielectric breakdown is an electrical process, its consequences are not limited to electrical aspects. The sudden discharge of stored electrical energy during breakdown can induce melting, vaporization, fracturing, and material displacement within the regolith. Thus, repeated breakdown events may continuously modify the physical structures of lunar regolith, a process known as dielectric breakdown weathering [52].
Dielectric weathering has the potential to progressively alter the physical characteristics of lunar regolith within PSRs. During breakdown, electrical discharge channels may propagate along mineralogical boundaries, creating fractures that weaken individual grains and promote fragmentation [82]. This suggests that repeated breakdown events may contribute to the production of fine grains with a high percentage of monomineralic grains. Breakdown-induced weathering may also increase the porosity of regolith by fragmenting the grains small enough to be dominated by the van der Waals force and facilitate their redistribution within the regolith [52, 85]. In addition, dielectric breakdown initiates vaporization of the material, and the vapor may condense onto surrounding grains thus contributing to localized reductions in albedo [52, 86].

Beyond these physical effects, weathering may also control the future electrical behavior of the regolith. Changes in grain size, porosity, microstructure, and mineral interfaces are expected to alter the dielectric properties of regolith, potentially modifying its susceptibility to future charging and dielectric breakdown events. As a result, dielectric breakdown may not only act as a weathering process but also contribute as a feedback mechanism that influences the long-term electrical evolution of the lunar regolith.

## 3 Observational Evidence and Case Studies

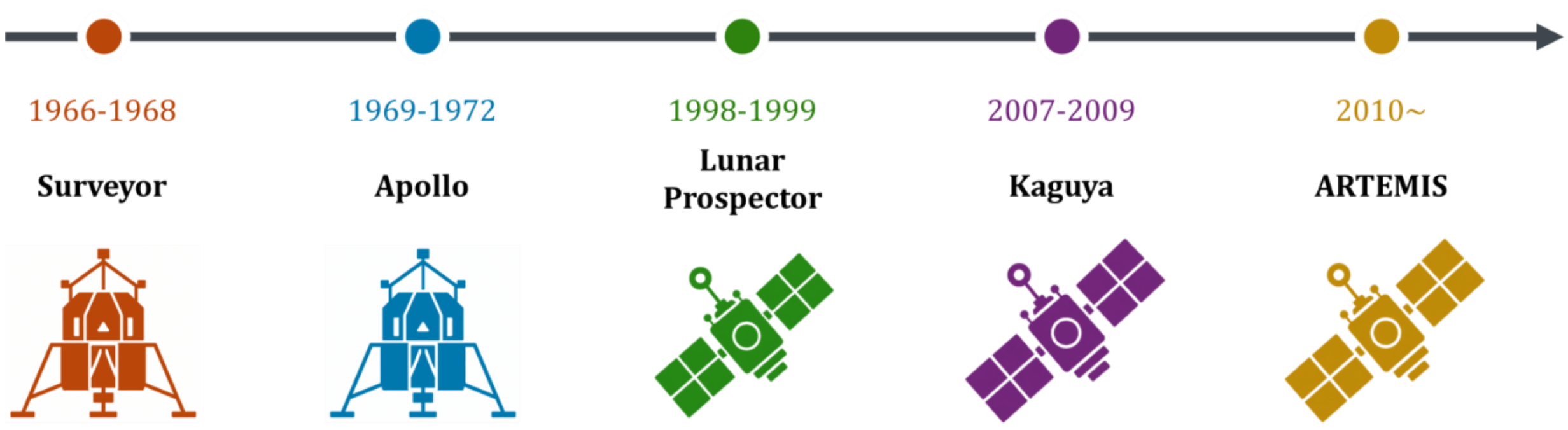

*Table 2. Summary of reported lunar surface potentials from in situ and orbital measurements across different plasma environments and solar illumination conditions.*

| Mission | | Surface Potential (V) | | | | | | Remark |
|---|---|---|---|---|---|---|---|---|
| | | Magnetotail | | | | Solar wind | | |
| | | Tail lobe | | Plasma Sheet | | | | |
| | | Day | Night | Day | Night | Day | Night | |
| Surveyor | | Observational Evidence Only | | | | | | In situ [87] |
| Apollo | 12, 15 | N/A | N/A | N/A | N/A | 10 | N/A | In situ [88, 89] |
| | 14 | > 200 | N/A | N/A | N/A | 10 | N/A | In situ [90] |
| Lunar Prospector | | < 20 | -150 ~ 0 | -1000 ~ 0 | -1000 ~ 0 | < 20 | -200 ~ 0 | Orbital [91] |
| Kaguya | | N/A | -500 | N/A | N/A | N/A | N/A | Orbital [92] |

| | | | | | | | |
|---|---|---|---|---|---|---|---|
| | | ~ -100 | | | | | |
| ARTEMIS | 15 ~ 25 | N/A | -450 ~ -180 | N/A | N/A | N/A | Orbital [93, 94] |

This chapter summarizes in situ and orbital measurements and observationally meaningful constraints on the lunar surface potential. Although reported values vary with measurement method, location, and environmental conditions, the results broadly confirm the expected behavior: the lunar surface charges to negative potentials on the nightside, where photoemission is absent and the surface collects ambient plasma electrons, and to positive potentials on the dayside. An exception occurs in the plasma sheet environment, where negative potentials have been reported even on the sunlit dayside, as described in Sections 3.3 and 3.5.
Not all missions reviewed in this chapter were designed to measure the lunar surface potential directly; however, missions that provided observationally meaningful constraints on the surface charging environment are included in this chapter. Among the missions reviewed, the Surveyor and Apollo missions conducted measurements from the lunar surface itself, while Lunar Prospector, Kaguya, and ARTEMIS acquired measurements from lunar orbit. Table 2 summarizes the representative surface potential values reported by each mission across different plasma environments and solar illumination conditions, where N/A indicates that measurements were not conducted or were not achievable under those conditions due to instrumental limitations. This comparative summary allows direct comparison of observational values obtained under varying environmental conditions.

### 3.1 Surveyor (1966 ~ 1968, NASA)

The earliest observational evidence associated with the lunar charging environment was obtained during the Surveyor program, which was conducted primarily to support the Apollo missions by characterizing the lunar surface and identifying suitable landing sites. Although seven robotic spacecraft were planned to be sent to the Moon, scientifically meaningful surface data were successfully returned by Surveyors 1, 3, 5, 6, and 7. Among these, Surveyors 5, 6, and 7 reported a faint glow extending along the western lunar horizon after local sunset, a phenomenon now commonly known as lunar horizon glow [87].

### 3.2 Apollo (1969 ~ 1972, NASA)

The Apollo program marked the first era of in situ measurements of the lunar electrostatic environment. Unlike the Surveyor horizon glow observations, which provided indirect visual evidence of dust lofting, several instruments were deployed to obtain quantitative measurements of charged particle fluxes and near-surface electric potentials. Among the experiments relevant to lunar surface charging, two provided data and constraints on the surface potential: the Suprathermal Ion Detector Experiment (SIDE) and the Charged Particle Lunar Environment Experiment (CPLEE).

#### 3.2.1 SIDE — Suprathermal Ion Detector Experiment (Apollo 12, 14, 15)

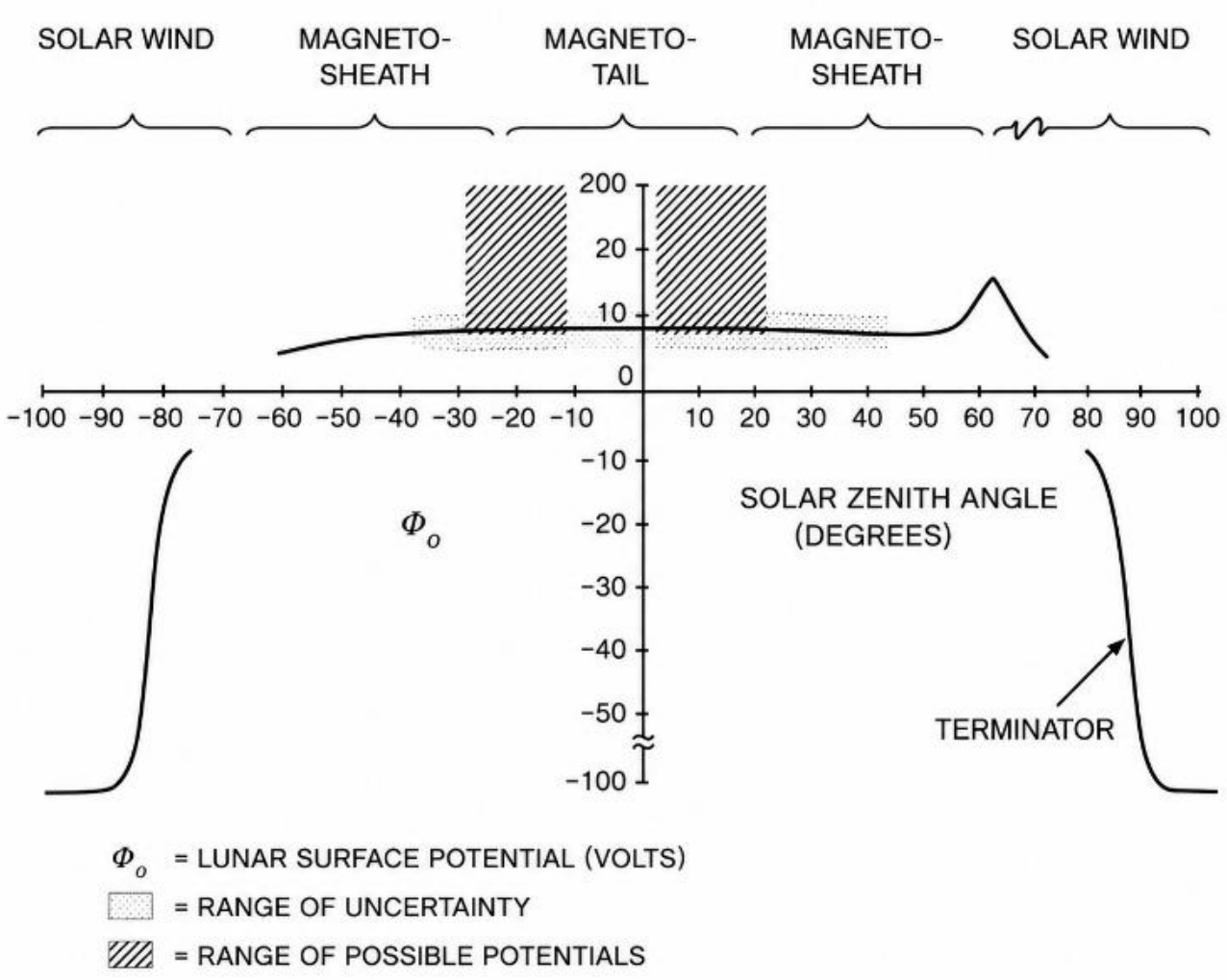

*Figure 17. Measured lunar surface potential as function of solar zenith angle [88].*

SIDE measured the energy spectra of positive ions near the lunar surface in the range of 10 eV to 3.5 keV. Surface potential was inferred through a resonance method: UV photons ionize neutral atoms just above the surface, generating photoions that are accelerated toward the detector by the local surface potential. The instrument steps a bias voltage applied at the detector entrance through discrete levels to selectively admit ions of different energies. When the bias is lower than the surface potential, the photoions carry insufficient energy to overcome the retarding field and are rejected; when the bias exceeds the surface potential, the admitted ion population broadens and the signal becomes diffuse. Only when the bias exactly matches the surface potential do all photoions arrive with a single characteristic energy eΦ, producing a sharply peaked flux in the spectrum. This resonance condition directly identifies the surface potential [88].

On the sunlit dayside at low solar zenith angles, SIDE consistently measured surface potentials on the order of +10 V [88, 89]. As the solar zenith angle increased toward the terminator, the measured potential decreased and crossed into negative values, reaching approximately −10 V in the region 20°–30° before the terminator [88]. Approaching the terminator itself, the inferred potential decreased further to −10 to −100 V [88]. Beyond the terminator, the resonance method lost sensitivity in full shadow, and SIDE therefore provided few quantitative constraints on nightside surface charging [88]. As shown in Figure 17, these measurements provided the first in situ evidence of a transition in lunar surface potential from positive values on the dayside toward negative values near the nightside, demonstrating the dependence of lunar surface potential on the ambient environment [91].

### 3.2.2 CPLEE — Charged Particle Lunar Environment Experiment (Apollo 14)

The CPLEE comprised two identical electrostatic analyzers with apertures located 26 cm above the lunar surface, measuring electrons and ions in the energy range between 40 eV and 50 keV. Analyzer A pointed toward the local lunar vertical, while analyzer B was directed 60° from vertical toward lunar west. Because the Sun subtends only a narrow angle, solar UV could enter only one analyzer at a time; therefore, simultaneous identical signals in both analyzers were interpreted as an isotropic electron population rather than UV contamination [90]. Numerical modeling further showed that the potential difference between the lunar surface and the 26 cm aperture height was only approximately 3 V. Thus, after applying this small correction, the fluxes measured at the analyzer height could be used to infer near-surface electrostatic conditions [90]. This interpretation was supported by lunar eclipse observations: when solar illumination was cut off, the measured electron flux dropped to nearly zero and recovered immediately after the surface re-entered sunlight. This confirmed that the detected electrons originated from surface photoemission rather than from ambient plasma electrons [90].

During magnetotail passage on the dayside, CPLEE observed isotropic returning electron fluxes in the 40–200 eV range, with no significant flux above this range [90]. These electrons were interpreted as photoelectrons emitted upward from the sunlit lunar surface, reflected back by the near-surface positive potential well, and accelerated downward toward the Moon. Because the detected return energy depends on the initial emission angle and analyzer geometry, the maximum observed energy provides a lower limit on the surface potential. Thus, the absence of returning photoelectrons above 200 eV indicates that the lunar surface potential is at least +200 V [90].

By contrast, under solar wind conditions and magnetosheath conditions, energetic ambient electrons overwhelmed the photoelectron signal, making it impossible to isolate the returning photoelectron population. Consequently, CPLEE-based surface potential measurements were only feasible during magnetotail intervals in which the ambient electrons were sufficiently low for the photoelectron signal to be distinguished [90]. These observations from CPLEE complemented SIDE measurements by showing that surface potentials are not solely determined by solar illumination. The inferred surface potentials of 200 V during magnetotail intervals indicate that variations in the ambient plasma environment are capable of modifying dayside charging, amplifying dayside surface potentials more than an order of magnitude compared to those typically observed under nominal solar wind conditions [91].

### 3.3 Lunar Prospector (1998 ~ 1999, NASA)

Lunar Prospector (LP) orbited the Moon at ~100 km altitude in a polar orbit. It carried a magnetometer (MAG) and an Electron Reflectometer (ER), a hemispherical electrostatic analyzer that measured electron differential energy flux across 32 energy channels (~7 eV to 20 keV). The ER rotated continuously, sweeping through all directions to produce a two-dimensional map of electron flux as a function of energy and pitch angle which is the angle between the electron velocity vector and the local magnetic field direction. The ER was originally designed to map lunar crustal magnetic fields by analyzing how incident electrons are magnetically reflected above the surface. However, the reflected electron energy distribution was found to be inconsistent with purely magnetic reflection and showed the influence of surface electric fields. Analysis of this reflected electron energy distribution enabled remote measurement of the lunar surface potential from orbit [95].

The measurement principle relies on the magnetic mirror effect. Electrons from the ambient plasma gyrate around and travel along magnetic field lines under the Lorentz force. Over localized lunar crustal magnetic anomalies, the magnetic field generally strengthens toward the surface because the crustal magnetic source lies below the spacecraft. As an electron moves from the weaker-field region at LP altitude toward the stronger-field region near the surface, conservation of the magnetic moment converts part of its field-parallel motion into field-perpendicular motion. For electrons with the same kinetic energy, the pitch angle determines this velocity partition: small pitch-angle electrons have a larger field-parallel component and reach the surface, whereas large pitch-angle electrons have a larger field-perpendicular component and are mirrored upward before impact. The critical pitch angle separating these two populations defines the loss cone boundary. In the purely magnetic case, this boundary is determined by the ratio of the magnetic field strength at LP altitude to that near the surface and is independent of electron energy [95].

When the lunar surface is negatively charged, the resulting electric field acts as an additional barrier against incoming electrons. Low-energy electrons lack sufficient kinetic energy to overcome this barrier and are reflected before reaching the lunar surface, causing them to be detected at pitch angles that would otherwise fall within the loss cone in the purely magnetic case. High-energy electrons, by contrast, carry enough energy to penetrate the electrostatic barrier and reach the lunar surface, producing a loss cone similar to the purely magnetic situation. The energy-dependent difference isolates the electric field contribution from the measured electron energy distribution, enabling derivation of the lunar surface potential. When high-energy electrons do reach the surface, they eject secondary electrons. These secondary electrons are emitted nearly anti-parallel to the incident direction and being low in energy, are accelerated by the potential difference between the lunar surface and LP. The energy gained during this acceleration directly reflects the potential drop, providing an independent measure of the lunar surface potential [95]. However, when the lunar surface potential is positive, it attracts both incident and surface-emitted electrons toward the lunar surface rather than repelling them. Because neither the energy-dependent loss cone signature nor the accelerated secondary electron beam is produced under positive surface-potential conditions, the technique is restricted to measurements of negative lunar surface potentials [91]. Note that the ER measures the potential difference between the surface and LP, not the absolute surface potential. Since LP itself charged negatively in shadow, initial estimates carried an unknown offset of ~35 V [95].

Under solar wind conditions, the nightside lunar wake showed surface potentials of −200 to 0 V, and dayside surface potentials below the measurement threshold of +20 V [91]. In the magnetotail, tail lobe conditions produced −150 to 0 V, while the plasma sheet drove potentials to −1000 to 0 V, including anomalous negative values observed even on the sunlit dayside during plasma sheet passage [91, 96]. During SEP events, nightside potentials reached up to −4.5 kV, with a nearly one-to-one correspondence with large solar proton events [17].

LP provided the first global-scale mapping of lunar surface charging, extending the spatially limited Apollo ground-truth measurements. Beyond this, LP revealed that the lunar surface can reach extreme negative potentials under disturbed plasma conditions, especially within the plasma sheet and during SEP events. These observations indicate that ambient plasma conditions can drive the lunar surface into more extreme charging conditions than those inferred from the Apollo missions, suggesting that extreme charging could be a key component of the lunar plasma environment. Also, LP provided the first quantitative measurements of the nightside and plasma-sheet charging on a global scale, which was previously inaccessible in the Apollo surface experiments. However, LP could only sample a given surface location at the time of orbital passage, precluding continuous monitoring of temporal variations at any fixed site. Furthermore, the ~100 km measurement altitude means that the technique senses the electrostatic potential structure at the electron reflection altitude rather than at the surface itself; complex potential profiles can cause the remotely sensed value to differ from the true surface potential. Consequently, some of the large negative potentials inferred by LP may reflect non-monotonic sheath structures above the surface rather than the surface potential itself, arguably leading to an overestimation of the magnitude of true surface potentials [93, 96].

### 3.4 Kaguya (2007 ~ 2009, JAXA)

Kaguya orbited the Moon carrying both electron and ion sensors, among which the Electron Spectrum Analyzer (ESA-S1), directed toward the lunar surface, and the Ion Energy Analyzer (IEA), directed away from the Moon, were used for surface potential determination. On the nightside in the terrestrial magnetotail lobes, cold ions in the vicinity of Kaguya were accelerated toward the negatively charged orbiter, and their measured energy directly gave the Kaguya potential. Simultaneously, electrons accelerated away from the more negatively charged lunar surface were detected by ESA-S1 with an energy corresponding to the potential difference between the surface and Kaguya. Adding the Kaguya potential derived from the ion measurement to this potential difference yielded the absolute surface potential [92]. The surface potential variations measured for both Kaguya and the Moon over time confirmed that both signatures originated from the same charge environment and supported the reliability of this method [92]. This technique was feasible in the magnetotail lobes, where the ambient plasma is cool enough for the accelerated ion peak and electron beam to be clearly resolved. In the plasma sheet or solar wind condition, energetic ions and electrons obscure both signatures.

On the nightside during the transition from the magnetotail lobe to the plasma sheet, the measured surface potential ranged from approximately −500 V to −100 V, with the magnitude correlating with the ambient electron temperature [92]. These values span the range between the LP-reported tail lobe potentials (−150 to 0 V) and plasma sheet potentials (−1000 to 0 V), consistent with the transitional plasma conditions under which the Kaguya observations were obtained [91].

Kaguya determined both the orbiter potential and the surface potential directly from the observed data, achieving the first purely observation-based determination of the absolute lunar surface potential from orbit. LP, by contrast, had no onboard means to measure its own electrostatic potential and could only determine the potential difference between the lunar surface and the spacecraft, so that the absolute surface potential had to be inferred by correcting for the LP potential using a spacecraft charging model, which introduced unavoidable uncertainty [91]. These measurements provided observational support for the existence of large negative surface potentials previously inferred from LP, though at magnitudes below the most extreme LP values. Furthermore, in the nightside magnetotail lobe environment, the observed correlation between surface potential and ambient electron temperature supported theoretical charging models in which energetic plasma electrons play a primary role in lunar nightside charging [92].

### 3.5 ARTEMIS (2010 ~ present, NASA)

ARTEMIS operates two probes (P1 and P2) in elliptical lunar orbits, measuring electron and ion energy spectra with the Electrostatic Analyzer (ESA) and directly determining the spacecraft floating potential with the Electric Field Instrument (EFI) [93]. EFI generates bias currents into spherical sensor electrodes at the tips of wire booms extending from the spacecraft body, maintaining the electrode potential close to the local plasma potential, and derives the spacecraft floating potential from the voltage difference between the electrode and the spacecraft body [93]. Two methods are used to measure the lunar surface potential. When the surface is at a negative potential, the same electron reflectometry technique as LP is applied, inverting the surface potential from the energy-dependent loss cone of incident electrons repelled by the surface electric field [93]. When the surface is at a positive potential, ion reflectometry is applied, inverting the surface potential from the energy-dependent loss cone of incident lobe ions repelled by the positive surface potential [97]. Unlike LP, the direct measurement of spacecraft potential by EFI allows the absolute lunar surface potential to be derived by combining the spacecraft-to-surface potential difference with the independently measured spacecraft potential [93].

In the dayside magnetotail lobe, ion reflectometry measurements indicate a positive lunar surface potential of approximately 15–25 V. This positive charging is attributed to the low ambient plasma density in the lobe, where the photoelectron emission current dominates over the incident electron current [97]. An empirical current balance model further predicts that the surface potential can increase by up to 100 V during strong solar flares [97]. However, electron reflectometry measurements in the same dayside lobe environment yield apparent negative potentials of approximately −30 to −60 V [93]. This discrepancy does not necessarily indicate that the surface itself is negatively charged in the dayside lobe. Instead, it reflects an interpretational limitation of electron reflectometry under non-monotonic potential conditions. A space-charge-induced potential well can form above the positively charged surface, causing electron reflectometry to respond to the potential minimum along the field line rather than to the actual surface potential [93, 96].

In the dayside plasma sheet, the charging behavior is more strongly affected by the denser and more energetic plasma environment. Electron reflectometry measurements reveal accelerated electron beams emanating from the surface and yield substantially more negative apparent potentials of approximately −180 to −450 V [93]. These values lie within the range reported earlier by Lunar Prospector under plasma sheet conditions (Section 3.3). The same interpretational caveat applies here as in the dayside lobe: the inverted value corresponds to the potential minimum along the field line rather than to the surface itself [93, 96].

In the magnetotail lobe nightside environment, surface potentials smaller in magnitude than predicted by current balance theory were occasionally observed by ARTEMIS [94]. The observed deviations were spatially correlated with the dawn-concentrated micrometeoroid flux measured independently by other instruments, identifying micrometeoroid impact-generated plasma as the likely source of the additional positive current partially neutralizing the negative surface charge [94]. This effect was detectable in the magnetotail lobe because, under solar wind or plasma sheet conditions, the surface potential is dominated by photoemission and ambient plasma currents whose magnitudes far exceed the micrometeoroid impact-generated current, rendering its contribution indistinguishable; in the low-density magnetotail lobe, where these dominant currents are greatly reduced, the micrometeoroid impact-generated current becomes a comparatively significant fraction of the total current balance and its effect on the local surface potential becomes observable [94].

The direct measurement of spacecraft potential by EFI eliminates the spacecraft potential uncertainty that limited LP. ARTEMIS provided the first orbital observational confirmation that non-monotonic potential structures exist above the dayside surface in the magnetotail lobe [93], and the first in situ evidence that micrometeoroid impact-generated plasma contributes measurably to the lunar surface potential [94], establishing it as an additional current source in the lunar charging current balance. Limitations include the variable measurement altitude associated with the elliptical orbit and the requirement that magnetic field lines intersect the lunar surface for both electron and ion reflectometry measurements, preventing continuous observations [93].

# 4 Discussion and Conclusions

The lunar surface potential is determined by the balance between the currents arriving at and leaving the surface, with contributions from photoelectron emission, ambient plasma electron and ion collection, and secondary electron emission [4, 6]. Because each of these responds to a different part of the environment, the potential differs with location and time. The time variation follows the diurnal cycle, the monthly passage through the terrestrial magnetotail, and the solar cycle [47, 49]. Regional differences arise from the composition and physical properties of the regolith, which are not uniform across the surface [35, 37]. The measurements compiled in Chapter 3 bear this out: reported values range from positive potentials of order tens of volts to negative potentials of kilovolt magnitude [88, 90, 91, 93, 98], and at a single location both the magnitude and the sign change as the surrounding environment changes [92, 94].

The current-balance framework describes an instantaneous equilibrium at the surface and does not follow charge that has penetrated into the material. Where the relaxation time is short this is immaterial. Within permanently shadowed regions the conductivity falls sharply at cryogenic temperatures and the relaxation time lengthens to a scale of weeks, comparable to the interval between SEP events [19, 41]. When charge arrives before the previous deposit has dissipated, accumulation rather than instantaneous balance governs the outcome, and because SEPs and GCRs reach millimeter and centimeter depths respectively [52, 54, 55], that accumulation occurs within a volume. Chapter 2 examines where this volumetric charging originates. Its central result is that energy deposition and charge retention are distinct outcomes. An incident particle may be stopped, backscattered, or transmitted, and it can deposit most of its kinetic energy while its own charge leaves the control volume [62]. A particle that is stopped and retains its charge contributes no net charge if the secondary electrons generated along its track escape. Escape is likewise not a single event: crossing the grain of origin, traversing a pore, and leaving a bulk control volume are physically distinct, and an electron reabsorbed by a neighboring grain produces inter-grain charge separation without changing the net charge of a volume containing both [62, 64]. A single interaction therefore cannot be reduced to one net charge value; it must be described as a spatially separated retained-charge distribution.

The microscopic input required by a macroscopic model follows from this. This memorandum defines it as the average signed retained charge per unit depth per incident particle, resolved by particle species and incident energy. Combined with the particle flux and energy spectrum, it yields a depth-dependent volumetric charge-source rate, which can then be coupled with charge continuity, conduction, dielectric relaxation, and Poisson's equation to describe the evolution of the subsurface charge density and internal electric field. Existing PSR charging models already supply this macroscopic framework, whether by treating the subsurface as a parallel-plate capacitor between two planar layers [19, 71] or by representing the currents and charge density as continuous functions of depth [66]. Neither derives its charge-source term from the retained-charge response per incident particle, and the CSDA range does not close the gap either, since it provides a characteristic transport scale while neglecting straggling, backscattering, transmission, and the spatial distribution of deposited energy [67, 68]. This connection between the microscopic interaction and the macroscopic charge source is the gap identified here.

The formulation rests on three assumptions whose validity conditions are not established: that grain- and pore-scale heterogeneity can be replaced by a depth-dependent ensemble average over a sufficiently large control volume; that the retained-charge responses of successive particles superpose approximately linearly; and that microscopic retained-charge formation completes faster than bulk conduction and dielectric relaxation. The minimum averaging scale, the charging level at which accumulated potential begins to alter secondary-electron escape and recollection, and the degree of timescale separation under cryogenic conditions all remain to be determined. These are the first targets for laboratory measurement and particle-resolved modeling.

Where the internal field exceeds the dielectric strength of the regolith, dielectric breakdown follows [52, 74]. PSR regolith favors this process: low conductivity and long relaxation allow the field to build, angular grain geometry enhances it at sharp asperities [78], and dielectric discontinuities at mineral inclusions and pores concentrate it further [79, 80, 81, 82, 83]. Breakdown in turn alters grain size, porosity, microstructure, and albedo through localized melting, vaporization, and fragmentation [52, 82, 85, 86], and these are the same properties that set the charging behavior and the breakdown threshold. Breakdown therefore acts as a feedback on subsequent charging rather than as weathering alone. The mechanisms invoked here are established for dielectric materials in general; their quantitative contribution under lunar conditions has not been determined.

Several elements of the problem are established: the current-balance framework accounts for the surface potentials reported by successive missions, the penetration depths of SEPs and GCRs and the charge relaxation timescales of cryogenic regolith are quantified, and macroscopic models of subsurface charging already exist. This memorandum has reviewed that context and has defined one structural gap explicitly: the connection between an individual incident particle and the volumetric charge source those models assume. It specifies the required input as the average signed retained charge per unit depth per incident particle and sets out the route from that quantity to a depth-dependent charge-source rate and the resulting internal field. Closing the gap requires quantities that are not yet constrained: the charge contributions of stopped, implanted, backscattered, and transmitted primaries, the fractions of generated electrons that recombine, are recollected or reabsorbed, become trapped, or escape, and the dependence of all of these on mineralogy, impact-glass content, grain and pore geometry, temperature, and pre-existing potential. Cryogenic high-vacuum irradiation of PSR-relevant materials, combined with particle-resolved modeling, is required to determine them.

Since Surveyor reported the first evidence of lunar charging in the form of horizon glow after local sunset [87], several missions have measured the lunar surface potential. The Apollo measurements are the most direct in that they were made at the surface itself [88, 89, 90], but they were obtained at a small number of equatorial sites over limited intervals, and each technique was valid only within a narrow range of illumination and plasma conditions. Orbital measurements from Lunar Prospector, Kaguya, and ARTEMIS remove this spatial restriction and provide global coverage [91, 92, 93, 94], but they observe from altitude rather than at the surface, so the surface potential is not measured directly but inverted through a chain of assumptions concerning the spacecraft potential, the magnetic connection to the surface, and the structure of the potential above it [91, 96]. No mission to date has measured beneath the regolith directly.

The Artemis program is directed toward the lunar south pole. The permanently shadowed regions there combine every accumulation condition discussed above. Without solar illumination there is no photoemission and no diurnal reset of the charging state, and at cryogenic temperatures the conductivity falls sharply so that the charge relaxation time lengthens to a scale comparable to a lunation [19, 41]. These are the conditions most favorable to charge persisting and accumulating at depth rather than dissipating, and existing models predict that in this environment the internal electric field can approach or exceed the dielectric strength of the regolith [19, 66]. That prediction cannot be verified without measuring the subsurface directly. The next step therefore requires an observational platform capable of measuring the surface potential together with a depth-resolved profile of the internal electric field and of following its variation in time. Only with such observations can the lunar electrostatic environment be described quantitatively from the surface through the subsurface, which is a prerequisite for the hazard assessment and dust-mitigation design required for a sustained polar presence.